\documentclass[
aps,
twocolumn,prx,
showpacs,preprintnumbers,nofootinbib,
amsmath,amssymb,
aps,
superscriptaddress]{revtex4-1}
\usepackage[breaklinks=true]{hyperref}
\usepackage{breakcites}
\usepackage{float}
\usepackage{graphicx}
\usepackage{dcolumn}
\usepackage{bm}
\usepackage{hyperref}
\usepackage{ulem} 
\usepackage{breakcites}

\newcommand{\be}{\begin{equation}}
\newcommand{\ee}{\end{equation}}
\newcommand{\bi}{\begin{itemize}}
\newcommand{\ei}{\end{itemize}}
\newcommand{\bea}{\begin{eqnarray}}
\newcommand{\eea}{\end{eqnarray}}

\usepackage{color}

\begin{document}

\title{Black-hole spectroscopy, the no-hair theorem and GW150914: Kerr vs. Occam.}


\affiliation{Instituto Galego de F\'{i}sica de Altas Enerx\'{i}as, Universidade de
Santiago de Compostela, 15782 Santiago de Compostela, Galicia, Spain}
\affiliation{Department of Physics, The Chinese University of Hong Kong, Shatin, N.T., Hong Kong}
\affiliation{School of Physics and Astronomy, Monash University, VIC 3800, Australia}
\affiliation{OzGrav: The ARC Centre of Excellence for Gravitational-Wave Discovery, Clayton, VIC 3800, Australia}

\author{Juan Calder\'on~Bustillo$^{1,2,3,4}$}\noaffiliation
\author{Paul D. Lasky$^\text{3,4}$}\noaffiliation
\author{Eric Thrane$^\text{3,4}$}\noaffiliation

\pacs{04.80.Nn, 04.25.dg, 04.25.D-, 04.30.-w} 

\begin{abstract}

The ``no-hair'' theorem states that astrophysical black holes are fully characterised by just two numbers: their mass and spin. The gravitational-wave emission from a perturbed black-hole consists of a superposition of damped sinusoids, known as \textit{quasi-normal modes}.  Quasi-normal modes are specified by three integers $(\ell,m,n)$:  the $(\ell, m)$ integers describe the angular properties and $(n)$ specifies the (over)tone. If the no-hair theorem holds, the frequencies and damping times of quasi-normal modes are determined uniquely by the mass and spin of the black hole, while phases and amplitudes depend on the particular perturbation. Current tests of the no-hair theorem, attempt to identify these modes in a semi-agnostic way, without imposing priors on the source of the perturbation. This is usually known as \textit{black-hole spectroscopy}. Applying this framework to GW150914, the measurement of the first overtone led to the confirmation of the theorem to $20\%$ level. We show, however, that such semi-agnostic tests cannot provide strong evidence in favour of the no-hair theorem, even for extremely loud signals, given the increasing number of overtones (and free parameters) needed to fit the data. This can be solved by imposing prior assumptions on the origin of the perturbed black hole that can further constrain the explored parameters: in particular, our knowledge that the ringdown is sourced by a binary black hole merger. Applying this strategy to GW150914 we find a natural log Bayes factor of $\sim 6.5$ in favour of the Kerr nature of its remnant, indicating that the hairy object hypothesis is disfavoured with $<1:600$ with respect to the Kerr black-hole one.



\end{abstract}
\maketitle


\section{\textbf{Introduction}}

After fifteen confirmed observations of binary black hole mergers \cite{GWTC1,GW190412,Abbott:2020uma,NSBH,GW190521D}, the gravitational-wave detectors Advanced LIGO \cite{AdvancedLIGO} and Virgo \cite{AdvancedVirgo} are regularly observing the strongest regime of gravity \cite{GWTC1_TGR}, granting access to its most fundamental properties. The no-hair theorem \cite{Carter1971,Israel1967} is a remarkable result of General Relativity \cite{Einstein:1916vd,Misner:1974qy} stating that black holes are simple objects which, omitting any electric charge, are fully characterised by their mass and spin. The newborn, highly distorted black hole left behind by a binary black hole provides a perfect arena to test this idea. The gravitational waves emitted by this black hole as it settles to its final state, known as \textit{ringdown} emission, consist of a superposition of damped sinusoids, known as \textit{quasi-normal modes}. The two polarizations $h_+$ and $h_\times$ of the gravitational wave strain can be expressed as \cite{Vishveshwara1970,Press1971,Teukolsky1973,1975}:

\begin{equation}
\begin{aligned}
     h(\theta, & \varphi; t)= h_+(t) - ih_{\times}(t) = \\ & \sum_{\ell,m,n} Y_{\ell,m}^{-2} (\theta,\varphi) A_{\ell m n} e^{-t/\tau_{\ell m n}} e^{(i(2\pi f_{\ell m n} t + \phi_{\ell m n}))}. 
    \end{aligned}
\label{eq:QNM}
\end{equation}
According to the no-hair theorem, the damping times and frequencies $(\tau_{\ell m n}, f_{\ell m n})$ of these modes are solely determined by the mass and spin of the black hole, while the amplitudes $A_{\ell m n}$ and relative phases $\phi_{\ell m n}$ depend on the initial conditions of the perturbation.
For instance, with the exception of the recent observation GW190521 \cite{GW190521D,HeadOnArxiv,Isobel_ecc,Gayathri,Boson}, for all current observations, such perturbations can be safely assumed to be sourced from the quasi-circular merger of two compact objects, in a quasi-circular inspiral ~\cite{Lim}. The angles $(\theta,\varphi)$ in Eq. \ref{eq:QNM} are the polar and azimuthal angles of a spherical coordinate system centered on the black hole, with $\theta=0$ aligned with the black hole spin \footnote{We note that Eq. 1 ignores the impact of retrograde modes, which should only be relevant for highly anti-aligned spins with respect to the orbital angular momentum.}.

Quasi-normal modes are commonly described by integers $(\ell, m, n)$. 
First, the angular properties of the emission are described by spheroidal-harmonic indices $(\ell, m)$.
Second, each angular $(\ell, m)$ set supports different ``tones'' described by the integer $n$. 
The $(\ell=2,m=2,n=0)$ mode is commonly referred to as the fundamental mode 
Last, for each $(\ell,m,n=0)$ mode there are an infinite number of \textit{overtones} with $n\geq1$. These have lower frequencies and faster damping times than their corresponding $n=0$ tones. It is frequently argued that the independent measurement of two quasi-normal modes would allow to test their consistency and, therefore, to test the no-hair theorem. This is the main goal of a research program commonly known as ``black-hole spectroscopy'' \cite{Dreyer2004,Berti2006,Berti2016,Bhagwat:2017tkm,Thrane:2017lqn,Carullo:2019flw,Giesler:2019uxc,Isi:2019aib,Bhagwat2020,Ota2020,Cabero2020,Xisco,Brito_modes, Carullo_Empirical_Tests,Gossan_Bayesian_Selection,Tiger_Ringdown}.

The feasibility of the above measurement has been widely discussed in the literature, especially during the last year, with two aspects receiving most of the attention.
First, Eq. \ref{eq:QNM} is only valid when the perturbation undergone by the black hole is weak enough that the black hole is in its linear regime. Consequently, much attention has been dedicated to finding a prescription for the instant when the final black hole enters such regime after its formation. On the one hand, starting the analysis too early in the evolution of the black hole would yield biased results. On the other, waiting too long would dramatically reduce the available signal power due to the exponentially decaying nature of the emission. Many studies have been devoted to this matter \cite{Cabero:2017avf, Thrane:2017lqn, Bhagwat:2017tkm,Carullo:2019flw,Giesler:2019uxc,Masha_NonLinearities,Carullo_Empirical_Tests}, leading to different prescriptions and attempts to test the no-hair theorem on available gravitational-wave data. Carullo et al. \cite{Carullo:2019flw} searched for multiple angular ringdown modes in the gravitational-wave signal GW150914 $h_{\ell m 0}$, starting their analysis $\approx 3$ms after the signal peak. They demonstrated the presence of a ringdown mode with $f\sim 234$Hz and $\tau \approx 3.9$ms, consistent with those estimated by the LIGO and Virgo Collaborations \cite{GW190514-TGR}. Next, they looked for the presence of a second angular mode, finding no evidence for it. More recently, Giesler et al. \cite{Giesler:2019uxc} obtained the remarkable result that the usage of overtones allows to observe the linear regime at the signal peak (when the gravitational-wave strain is maximal), resulting in a great increase of the available signal power. Applying this idea to GW150914, Isi et al. \cite{Isi:2019aib} reported the measurement of the $(2,2,1)$ overtone, and its consistency with the no-hair hypothesis to a $20 \%$ level.

The work by Giesler at al. \cite{Giesler:2019uxc} and Isi et. al. \cite{Isi:2019aib}, has triggered the second main area of discussion \cite{Giesler:2019uxc,Ota2020,Bhagwat2020,Ota2020}: whether overtones of the fundamental mode, i.e., modes with $(\ell=2,m=2,n>0)$ or higher angular modes, i.e., those with $(\ell,m,n=0)$ and $(\ell,m) \neq (2,2)$, are the best candidates for the observation of a secondary mode, in addition to the fundamental $(2,2,0)$. The current prevailing view is that for remnants of nearly equal-mass binaries, overtones provide the best avenue. The reason is that while overtones always damp quickly, higher angular modes are highly suppressed by the symmetries of the source \cite{Ota2020,Blanchet2014,Bustillo:2016gid}. In contrast, for asymmetric binaries, angular modes are strongly triggered so that the $(3,3,0)$ mode, usually the strongest of them, is a better candidate. The reason is that the frequency of this mode differs more from that of the $(2,2,0)$ than that of the $(2,2,n \neq 0)$ overtones, making it easier to resolve \cite{Bhagwat2020,Xisco}. 

We note, however, that the above works \cite{Bhagwat2020,Xisco,Ota2020} do not involve full Bayesian model selection (see e.g. \cite{Gossan_Bayesian_Selection}), but rely on distinguishability criteria based on the Fisher Matrix formalism \cite{Lindblom:2008cm}, which neglects the impact of the size of the searched parameter space. Since testing the no-hair theorem ultimately involves performing model selection between at least two models (one satisfying the theorem and one violating it), the size of the parameter space is a crucial ingredient of model selection that cannot be ignored. In this sense, while the realisation that using overtones allow to extend the analysis up to the signal peak is a major advance, \cite{Giesler:2019uxc} also reveals that up to eight overtones may be needed to correctly fit the data for sufficiently loud signals. This means that one may need to use $2+2\times 8 = 18$ intrinsic parameters to describe the signal while the initial configuration has, at most eight. 
It should give us all pause that this framework seeks to model the remnant of a binary black hole merger using more physical degrees of freedom than those of the parent binary!\footnote{We note that recently, Jim\'{e}nez et. al. \cite{Xisco} have started to build overtone models parametrised on the binary parameters, which will eventually solve this issue. However, up to date, this only considers the first overtone.} 
More importantly, this greatly affects Bayesian model selection due to the increase of the parameter space, reflected in the \textit{Occam factor} (see Appendix II).

In this work, we use full Bayesian inference to study different frameworks for testing the no-hair theorem.
We critically assess the prevailing black-hole spectroscopy paradigm, which seeks to measure individual ringdown tones \cite{Isi:2019aib,Ota2020}. By performing model selection on numerically simulated signals consistent with GW150914 \cite{SXS}, we find that such test \textit{cannot} provide conclusive evidence that final object is a Kerr black hole, even when the signal-to-noise ratio (SNR) reaches $\rho=100$, around 8 times louder than for GW150914. The reason is that the increasing number of parameters needed to fit the data as the signal loudness grows, together with the strong constraints imposed on the properties of the quasi-normal modes by the no-hair theorem, lead to an important increase of the Occam factor. As a consequence,``hairy'' models including fewer modes not subject to the no-hair theorem constraints, cannot be confidently ruled out.

The solution, we argue, is to take into account our prior knowledge that the ringdowns observed by LIGO--Virgo are sourced by binary mergers  (see \cite{Brito_modes} for a similar strategy involving angular modes).
To that end, we model the ringdown signal using the ringdown part of gravitational waveform approximants for binary mergers tuned using numerical relativity. These waveforms encapsulate the information of all possible overtones, effectively imposing appropriate priors on their amplitudes and phases parametrised in terms of the eight parameters of the parent binary. The reduced parameter space of this ``binary'' model reduces the Occam factor, and allows for a much better determination of the Kerr nature of the final object. 

We carry out analyses of GW150914 using three different models, each consisting of a different implementation of Eq. (1): (a) A ``hair'' model in which all parameters run freely; (b) a ``Kerr'' model with $(\tau_{\ell,m},f_{\ell,})$ fixed by the mass and spin of the black hole; and (c) a binary black hole (``BBH'') model with all of the previous parameters fixed by the eight parameters of the parent binary. 
Using the BBH framework, we find a natural log Bayes factor $\log{\cal{B}}
^{\text{Kerr-BBH}}_{\text{Hair}} \sim 6.5$, leaving 1 in $\sim 600$ chances that the no-hair theorem may be violated. Using the spectroscopy framework, we obtain $\log{\cal{B}}
^{\text{Kerr}}_{\text{Hair}} \sim 1$, consistent with \cite{Isi:2019aib}. In addition, while our posterior distribution for the amplitude of the first overtone completely rules out zero, we only find $\log{\cal{B}} \sim 1.5$ in its favour, indicating a mild preference for the presence of such mode.

The rest of this paper is organised as follows. Section \ref{sec:models} describes the three signal models for ringdown emission considered in this study.
In Section \ref{sec:PE} we describe our analysis set-up on simulated signals and on the gravitational-wave signal GW150914. In Section IV, we report our results and we close our work with some final remarks.


\section{Signal models}
\label{sec:models}
We consider three different waveform models: Hair, Kerr, and BBH.
The three models are nested so that binary black hole is a sub-model of Kerr, which is a sub-model of Hair.
In each case, we approximate the waveform using the dominant quadrupole $(\ell,m)=(2,\pm 2)$ modes. 
We project the complex strain signal onto the Advanced LIGO detectors as:
\begin{equation}
\begin{aligned}
 h^{D}(t) = F_+ h_+ + F_\times h_\times,
\end{aligned}
\end{equation}
with 
\begin{equation}
\begin{aligned}
 h_+ - ih_\times =  Y_{(2,2)}(\iota,\varphi) h_{(2,2)}^{-2} + Y_{(2,-2)}^{-2}(\iota,\varphi) h_{(2,-2)}. 
\end{aligned}
\end{equation}
Above, $F_{+,\times}$ denote the antenna patterns of the detectors, which depend on the sky-location of the source and the polarisation angle, while the angles $(\iota,\varphi)$ denote the polar and azimuthal angle describing the location of the observer around the source in an spherical coordinate system centered at the source center-of-mass, with $\iota=0$ denoting the direction of the final spin.

\subsection{Hair model}
Our first and most general model consists of a superposition of $N$ damped sinusoids given by
\begin{equation}
\begin{aligned}
    & & h^{Hair}_{(2,\pm 2)}=\sum_{n=0,N} A_n e^{-t/\tau_{n}} e^{i\pm (2\pi f_{n} t + \phi_n))} .
    \end{aligned}
\end{equation}
Here, all $\{A_n,\phi_n,f_n,\tau_n\}$ parameters vary freely, effectively accounting for a wide variety of possible deviations from the no-hair theorem. 
To reduce the computational cost, and to avoid double mode counting, we impose the following constraint: $A_n < A_{n+1}$, $\tau_n > \tau_{n+1}$, $f_n > f_{n+1}$. This choice is motivated by the fact that, for the case of Kerr black holes, tones with larger $n$ are associated with smaller damping times, lower frequencies, and larger amplitudes than the fundamental tone. This way, our ``hair'' model can be understood as a ``hairy'' overtone model. A model with N tones will have $4N$ degrees of freedom. 
Evidence for the hair model (when compared to the Kerr model below) would suggest a source that is inconsistent with a general relativistic black hole.

\subsection{Kerr model}
Our second model---a sub-model of the hair model---requires that the frequencies and damping times of each tone are consistent with emission from a Kerr black hole. Thus, the frequencies and damping times are all functions of the final mass and spin $(M_f,a_f)$, while the amplitudes and phases are unconstrained:
\begin{equation}
\begin{aligned}
    & h^{Kerr}_{(2,\pm 2)}=\sum_{n=0,N} A_n e^{-t/\tau_{(2 2 n)}} e^{i\pm (2\pi f_{(2 2 n)} t + \phi_n))} . \\
    \end{aligned}
\end{equation}
Here, $f_{(2,2,n)}$ and $\tau_{(2,2,n)}$ denote the frequency and damping times of the $n$-th overtone of the $(2,2,0)$ mode. We stress that the addition of one overtone implies the addition of two extra parameters $(A_n,f_n)$ so that a model with N modes i.e., N-1 overtones, has $2N+2$ intrinsic degrees of freedom. 
This model is designed to make minimal assumptions about how a ringing black hole is perturbed.

\subsection{Binary black hole model}
Our third model---a sub-model of the Kerr model---requires that the amplitude and phase of each tone is consistent with excitation from a binary black hole merger.
 While currently there is no waveform model that explicitly provides the amplitudes and phases of the ringdown modes as a function of the binary parameters for spinning binaries \footnote{While this work was being performed, Jim\'{e}nez et. al., released a model for the remnant of non-spinning BBHs parametrising both the fundamental $(2,2,0)$ mode and its first overtone as a function of the binary parameters \cite{Xisco}. We will discuss this model later on.}, these amplitudes and phases are implicitly encapsulated in the post-peak of full inspiral-merger-ringdown models like \cite{Hannam:2013oca,Khan:2015jqa,Husa:2007hp,Cotesta:2018fcv,Blackman:2017pcm}. Since the post-merger of these waveforms is fitted to full numerical relativity simulations, these naturally include all the overtones, effectively allowing us to place priors on $\{A_n,\phi_n,f_n,\tau_n\}$ via the binary parameters. In this work, we use the phenomenological model for precessing binaries known as \texttt{IMRPhenomPv2} \cite{Hannam:2013oca}.
 




\begin{table}[t!]
\begin{ruledtabular}
\begin{tabular}{l l l  l}
 & $\rho=100$ &  $\rho=15$ & GW150914\\
\hline
\rule{0pt}{3ex}%
Hair, 1 mode & 4728 & 92.15 & 92.11 \\
\rule{0pt}{3ex}%
Hair, 2 modes  & 4963 & 89.03 & 93.88  \\
\rule{0pt}{3ex}%
Hair, 3 modes  & 4957 & 85.91  & 90.70  \\
\rule{0pt}{3ex}%
Hair, 4 modes  & 4954 & - -  & - -\\ 
\hline
\rule{0pt}{3ex}%

Kerr, 0 overtones & 4728 & 92.41  & 92.99  \\
\rule{0pt}{3ex}%
Kerr, 1 overtone & 4950  & 91.87 & 94.54 \\
\rule{0pt}{3ex}%
Kerr, 2 overtones & 4965 & 90.98 & 91.90  \\ 
\rule{0pt}{3ex}%
Kerr, 3 overtones & 4964 & 90.50 & - - \\
\hline
\rule{0pt}{3ex}%
Kerr BBH, Non-spin & 4971 & 97.03 & 100.56  \\
\rule{0pt}{3ex}%
Kerr BBH, Aligned spins & 4971 & 96.69 & 100.71  \\
\rule{0pt}{3ex}%
Kerr BBH, Precessing spins & 4971 & 96.45 & 100.94  \\
\hline

\end{tabular}
\end{ruledtabular}
\caption{Log Bayes factors for GW150914 and the numerical simulation SXS:BBH:0305 scaled to optimal SNRs of 15 and 100, when analysed with different ringdown models.}
\label{tab:parameters}
\end{table}

\begin{figure*}
\includegraphics[width=0.48\textwidth]{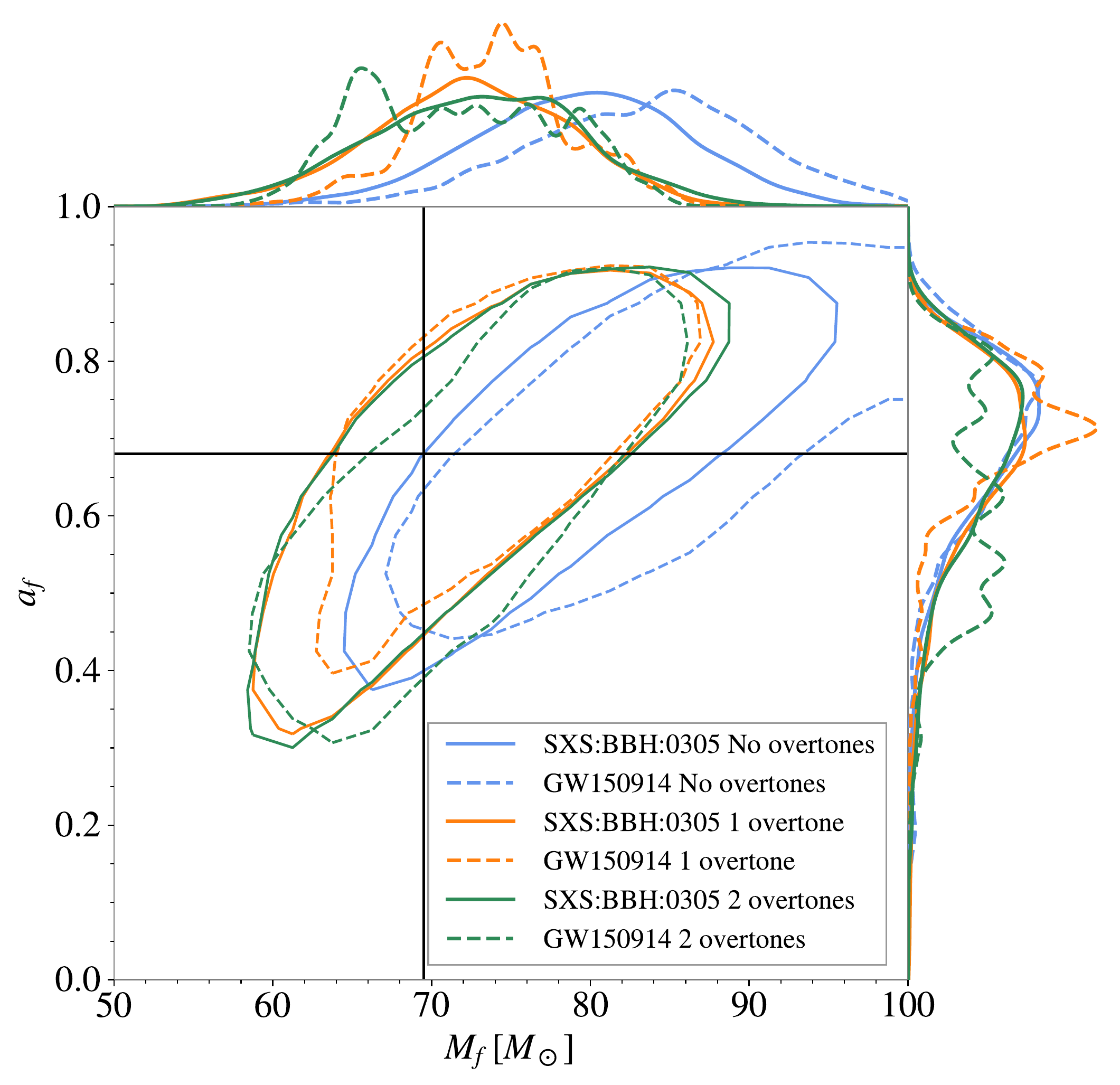}
\includegraphics[width=0.48\textwidth]{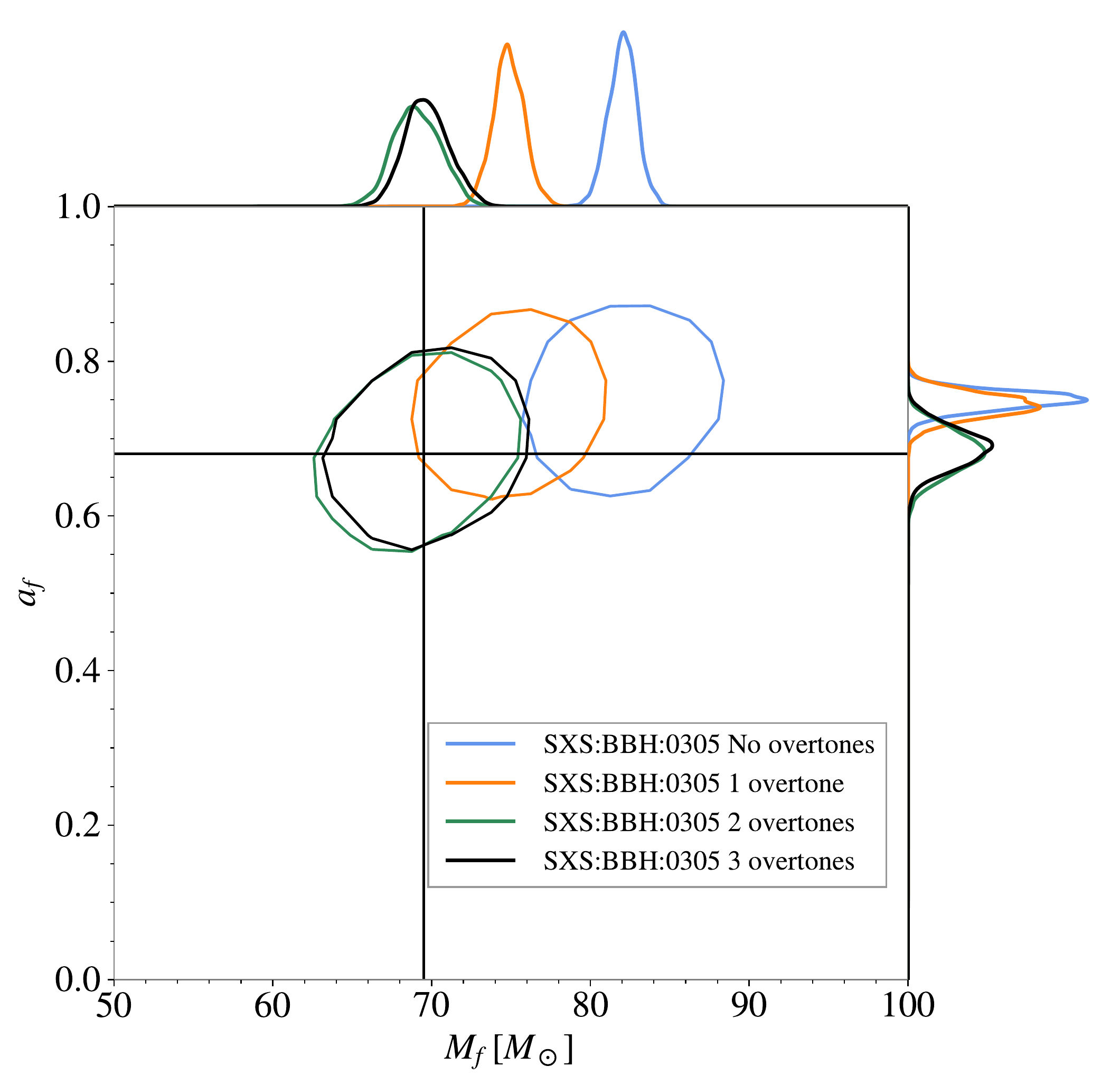}
\caption{\textbf{Left}: Final mass and spin recovery for GW150914 (dashed) and a consistent simulation with optimal SNR of $\rho_{opt}=15$ (solid) using 0,1 and 2 overtones (blue, orange, green). The contours obtained for both real and simulated data are in wide agreement. \textbf{Right}: final mass and spin estimates for the same injection as in the left panel, scaled to an SNR of 100. Two overtones are needed to obtain non-biased estimates. For our injection analysis, we consider a single Advanced LIGO detector implementing the noise curve obtained by BayesWave around the time of GW150914. The intersection of the black lines denotes the true values for our injection.} 
\label{fig:afmf}
\end{figure*}

\begin{figure*}
\includegraphics[width=0.49\textwidth]{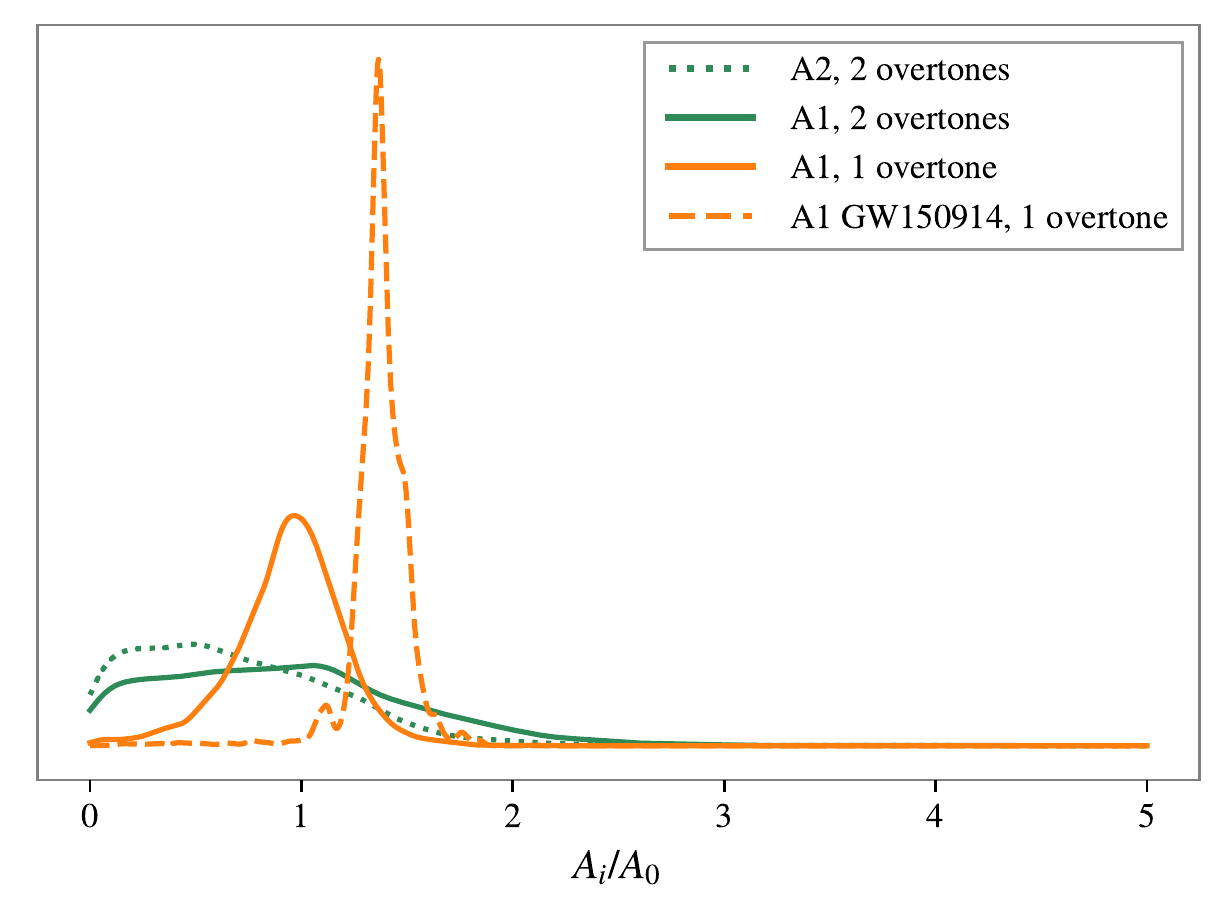}
\includegraphics[width=0.49\textwidth]{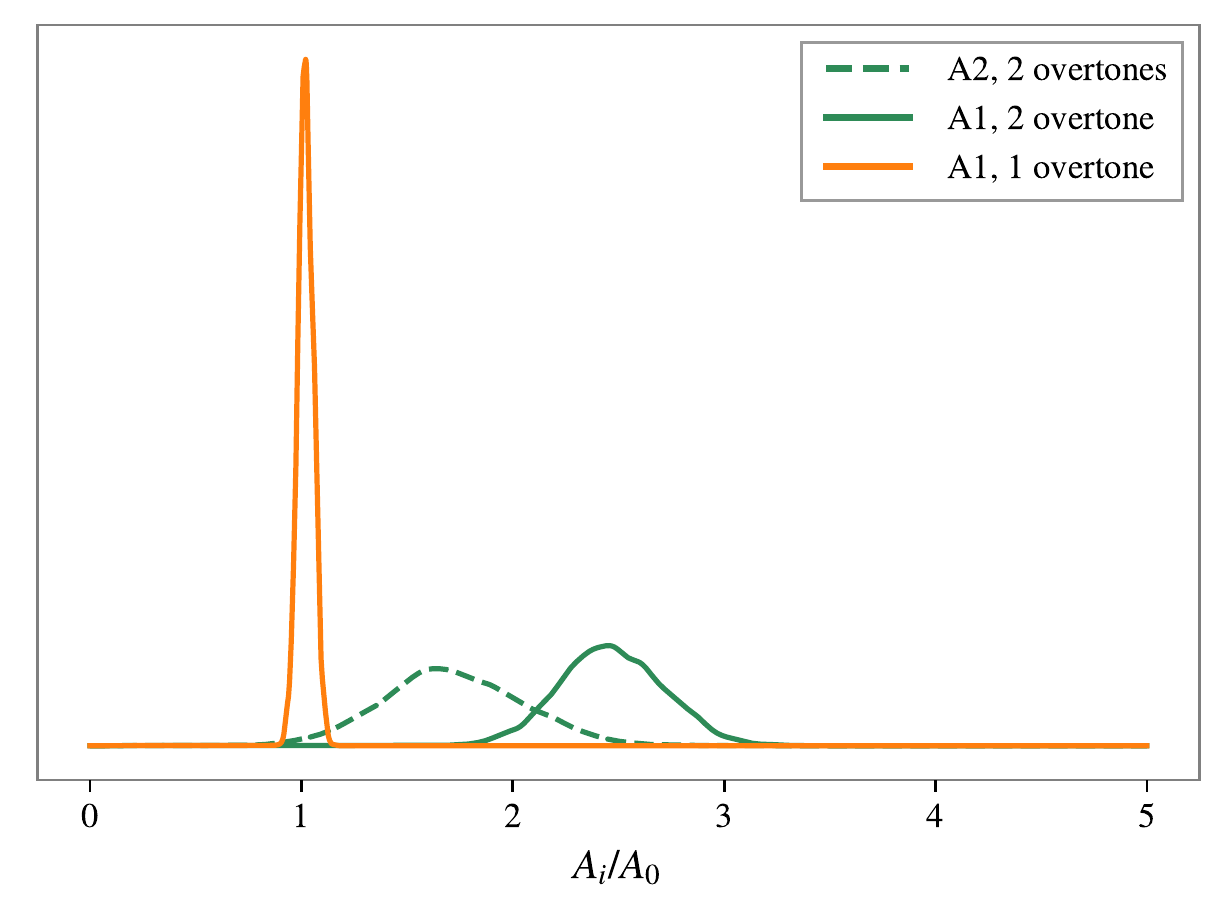}
\includegraphics[width=0.49\textwidth]{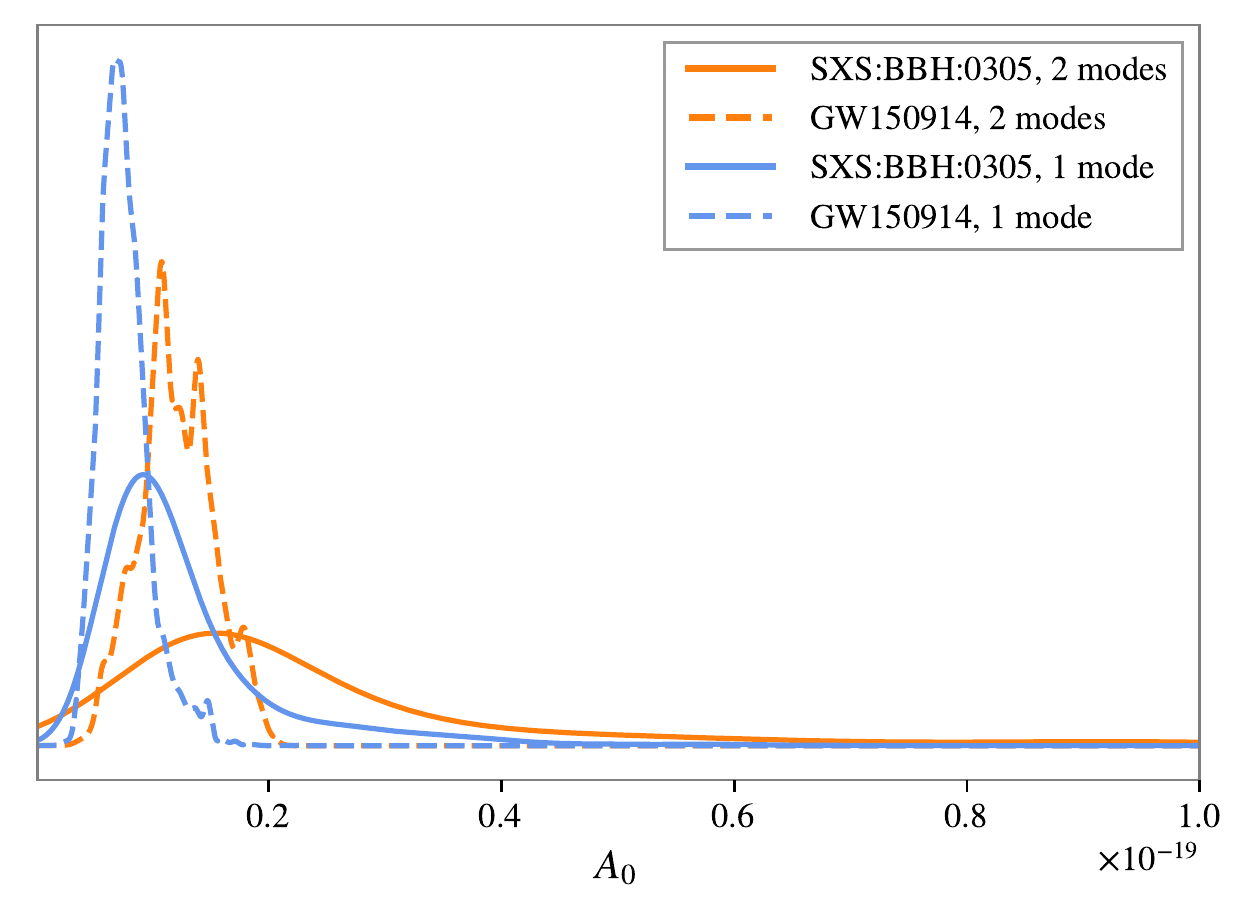}
\includegraphics[width=0.49\textwidth]{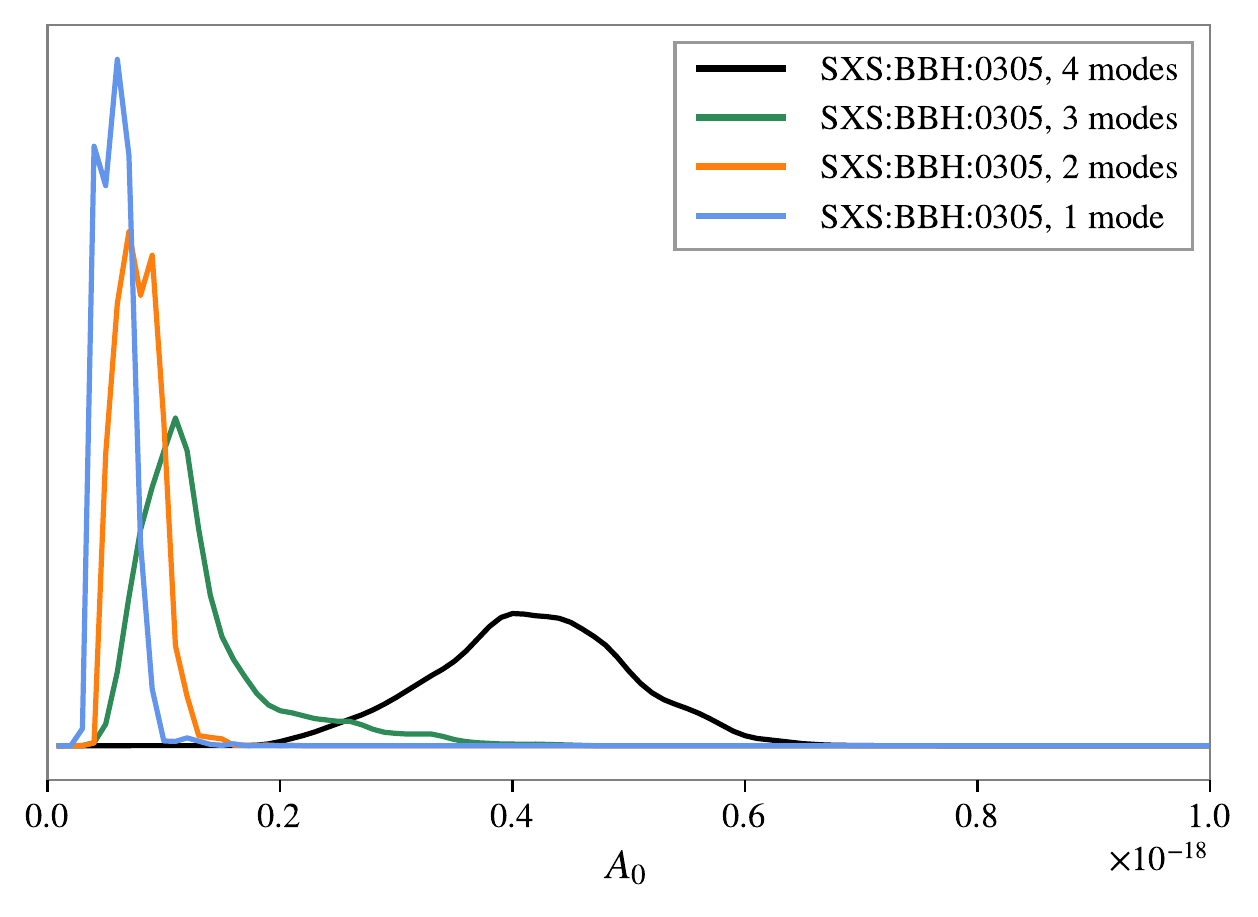}
\caption{\textbf{Overtone amplitudes for a GW150914 and a consistent simulation. Top:} Relative amplitudes $A_i/A_0$ of the overtones with respect to the fundamental tone. \textbf{Bottom}: Amplitude of the fundamental mode. The optimal SNR of the injections is $\rho=15$ in the left panels and $\rho=100$ on the right ones.} 
\label{fig:overtones}
\end{figure*}

\section{\textbf{Analysis of simulated signals}}
\label{sec:PE}

Our study consists of two steps: the analysis of simulated signals and that of real data from GW150914. First, we analyse a numerically simulated signal injected in zero-noise, with source parameters consistent with those of GW150914. We recover our injection with our Kerr and Hair models, varying the number of tones. We study the parameters recovered by each model and compare their Bayes Factors to asses the level to which the ``no-hair'' theorem and the presence of multiple modes can be tested. We do this analysis for a case where the signal loudness is consistent with that of GW150914: a post-peak optimal SNR of $15$; and again with an extreme-case with optimal SNR of $100$.

\subsection{\textbf{Setup}}

We consider a single Advanced LIGO detector with a noise power spectral density consistent with that obtained for GW150914 in \cite{GW150914_DATA,LVC_GW150914_noise_studies} using the BayesWave algorithm \cite{Cornish2015}. The number of resolvable overtones in the ringdown signal depends on the signal loudness, which is characterised by the optimal signal-to-noise ratio of the signal \cite{Finn1992,Cutler:1994ys}, given by:
\begin{equation}
   \rho_{opt}=(h|h)^{1/2}, 
\end{equation}
where 
\begin{equation}
   (a|b) = 4 \Re \int_{0}^{\infty} \frac{\tilde{a}(f)\tilde{b}^{*}(f)}{S_n(f)}df,
\end{equation}
denotes the inner product, and $\tilde{a}(f)$ denotes the Fourier transform of $a(t)$. We perform our analysis for optimal SNRs of $\rho_{opt}=15$, consistent with the post-peak SNR of GW150914, and $\rho_{opt}=100$. We implement a frequency domain likelihood given by \cite{Cutler:1994ys}:
\begin{equation}
 \log {\cal{L}}(\Theta|d) \propto -\frac{(d-h(\Theta)|d-h(\Theta))}{2}.
\end{equation}
We fix the sky-location and polarisation angle to the true ones, setting standard priors on all the other extrinsic parameters. 

We inject in zero-noise the numerical relativity binary black hole simulation \texttt{SXS:BBH:0305} waveform \cite{SXS}, with parameters consistent with GW150914; used in \cite{Giesler:2019uxc} (see Table I). To isolate the post-merger signal, we set to zero the data prior to the peak of the amplitude of the $(2,2)$ mode. We choose for the source to be face-on, to minimise the effect of higher-order angular modes \cite{Varma:2014jxa,Bustillo:2016gid,Pang:2018hjb,H0HM}.

We perform our parameter estimation runs using the code Bilby \cite{Ashton:2018jfp,RomeroShaw_bilby} and sample the parameter space using the CPNest sampler \cite{CPNest}.

\subsection{Real data analysis}

We analyse the post-peak portion of the signal GW150914, which we obtain from the publicly available Gravitational-wave Open Science Center \cite{GWOSC}. In accordance with \cite{Isi:2019aib}, we consider the peak to happen at the GPS time $t_\text{GPS} = 1126259462.423$ and impose a time delay of $6.9$ ms between the Livingston and Hanford detectors. Doing this kind of analysis presents challenges. In gravitational-wave data analysis, it is common to apply a window that smoothly sets the data to zero at the ends of the data segment to avoid spectral leakage, e.g., from noise lines. However, given the exponentially decaying nature of the signal after $t_\text{GPS}$, the application of a window to a data segment starting at $t_\text{GPS}$ would lead to a dramatic loss of signal power. On the other hand, starting the data segment prior to $t_\text{GPS}$ would lead to power leakages from the pre-peak portion that would affect our analysis.

A common solution to this is the implementation of a time-domain likelihood in which the power-spectral density $S_{n}(f)$ is replaced by a covariance matrix describing the covariance between the noise at different times \cite{Carullo:2019flw}. Here, we employ a different solution. We define a data segment of 4 seconds duration centered on the beginning of the ringdown: $t_\text{GPS} = 1126259462.423$. Next, we replace the pre-ringdown data with two seconds of representative noise data starting at $t_\text{GPS}+15$s. 
This method is designed to isolate the ringdown signal from the inspiral while allowing us to window the data at the ends of the segments without losing any post-peak signal power. Assuming that the replacement data has no glitches in it, and that the underlying gaussian stochastic
process describing the noise fluctuations is weakly-sense stationary
\cite{LVC_DA_guide}, as verified in \cite{LVC_GW150914_noise_studies} during the interval $t_\text{GPS}-2$s and $t_\text{GPS}+17$s, we can then use the PSD computed in the usual way.\footnote{PSDs are commonly constructed using longer data segments, so that such assumption is commonly made.}
As we show below, our results on both simulated and real data are broadly consistent. This makes us confident that our approach is sensible, at least for the analysis investigated here. 

\section{\textbf{Results}} 
In this section, we first study the parameters recovered by each of our models when analysing both our numerical simulation and GW150914. Then we report our model selection results. 

\subsection{Final mass and spin and overtone recovery}
The solid contours in Figure~\ref{fig:afmf} show the 2D $90\%$ credible intervals for the final mass and spin of the NR simulation scaled to an SNR of 15 (left) and 100 (right). In addition, the dashed contours in the left panel show the same intervals for GW150914. Figure~\ref{fig:overtones} shows the corresponding posterior distributions for the amplitude of the fundamental mode (bottom) and the relative amplitude of the overtones (top).
In agreement with previous studies \cite{Giesler:2019uxc,Isi:2019aib}, we find using solely one tone (omitting overtones) leads to biased parameter estimation. For a low SNR of 15, adding the first overtone is enough to correct this bias. Accordingly, we obtain a posterior distribution for the amplitude of the $n=1$ overtone that peaks away from zero. However, we find that the addition of this overtone produces a modest increase in the SNR $\delta \rho \sim 0.4$ while adding two extra parameters. This incurs an Occam penalty, which makes the Bayes Factor \textit{decrease}, yielding $\log{\cal{B}}^{n\leq 1}_{n=0} \approx 0.5$. Thus, from a Bayesian point of view, there is no evidence for a first overtone. Note that we obtain this result despite the fact that the recovered amplitude of the fist overtone is inconsistent with zero (see Fig.2, left upper panel). In particular, we obtain a $66\%$ interval for the amplitude of the first overtone of $A_1= 1.70^{+1.87}_{-0.69}\times 10^{-20}$ that excludes $A_1=0$ with more than $3\sigma$. 

The dashed contours in the left panel of Figure \ref{fig:afmf}, represent the $90\%$ credible intervals for the final mass and spin obtained for GW150914, which show wide agreement with those obtained for the NR simulation. In addition, similar results are shown in the rightmost column of Table I. Once again, while the addition of the first overtone yields parameter estimates consistent with the true values, and despite the fact that we obtain $A_1 = 1.62^{+0.81}_{-0.71} \times 10
^{-20}$ at the $66\%$ credible level, we obtain $\log{\cal{B}}^{n\leq 1}_{n=0} \approx 1.5$, indicative of a weak evidence for the first overtone. We note that this result is quantitatively consistent with that obtained by the LIGO and Virgo collaborations \cite{O3a_TGR} after the release of this work, using a time-domain analysis  \cite{Carullo:2019flw}.

When we increase the SNR of the injection to 100, the amplitude posterior of both the first and the second overtone clearly exclude zero (see Figure \ref{fig:overtones}, top-right panel). Moreover, the inclusion of these two overtones is needed to obtain non-biased posterior distributions. For this case, we obtain  $\log{\cal{B}}^{n\leq 1}_{n=0} \approx 222$ and $\log{\cal{B}}^{n\leq 2}_{n\leq 1} \approx 15$, yielding clear evidence for the presence of at least two overtones. When a third overtone is added, the log Bayes factor comparing the model with $N$ overtones to $0$ tones $\log{{\cal{B}}^{n \leq N}_0}$ starts to decrease.
The fit is slightly improved, but not enough to compensate for the increase in parameter space. This leads to $\log{\cal{B}}^{n\leq 3}_{n\leq 2} \approx -1$ and indicates that a larger signal loudness would be needed to confirm the presence of a third overtone.

\begin{figure*}
\includegraphics[width=0.32\textwidth]{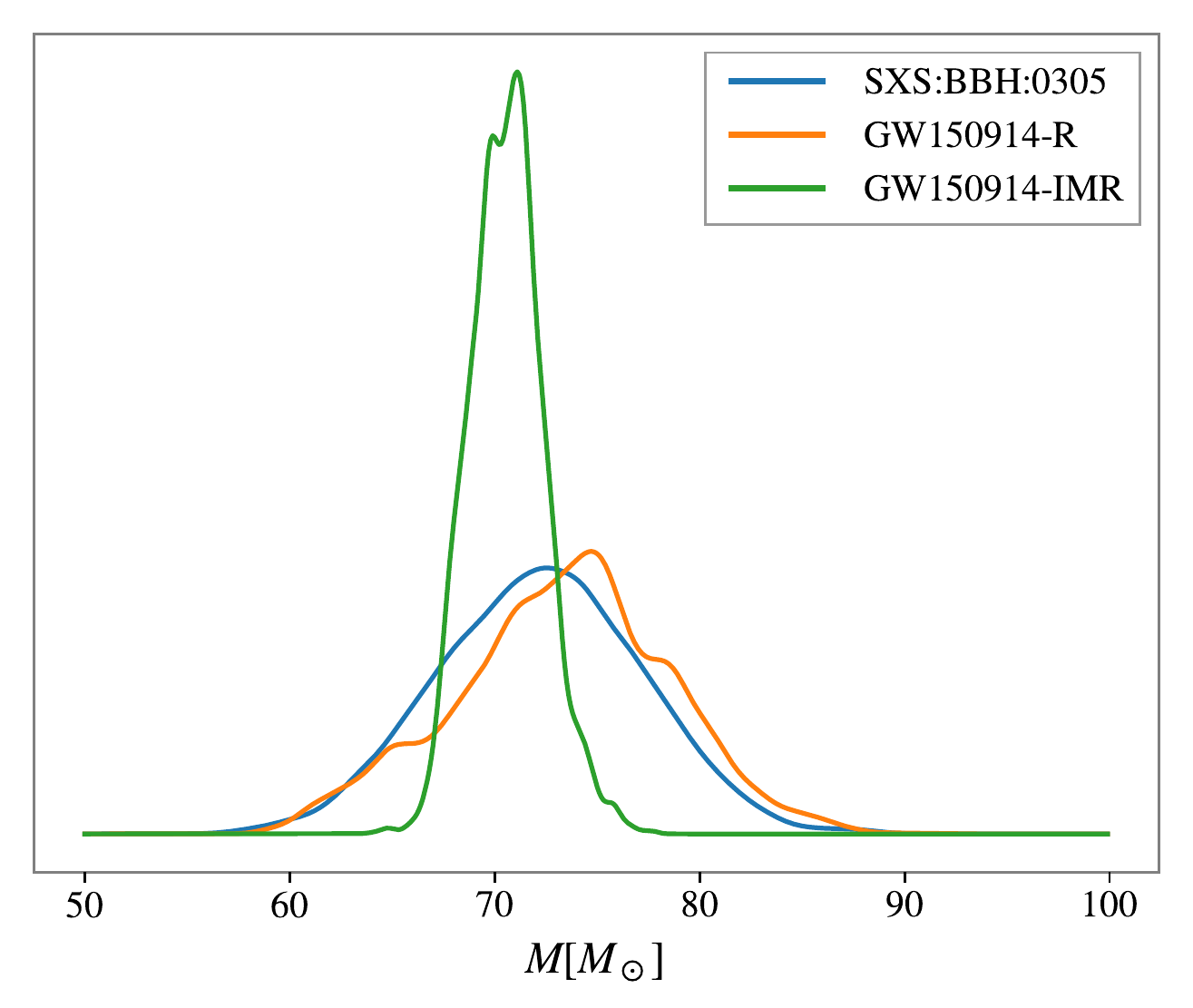}
\includegraphics[width=0.32\textwidth]{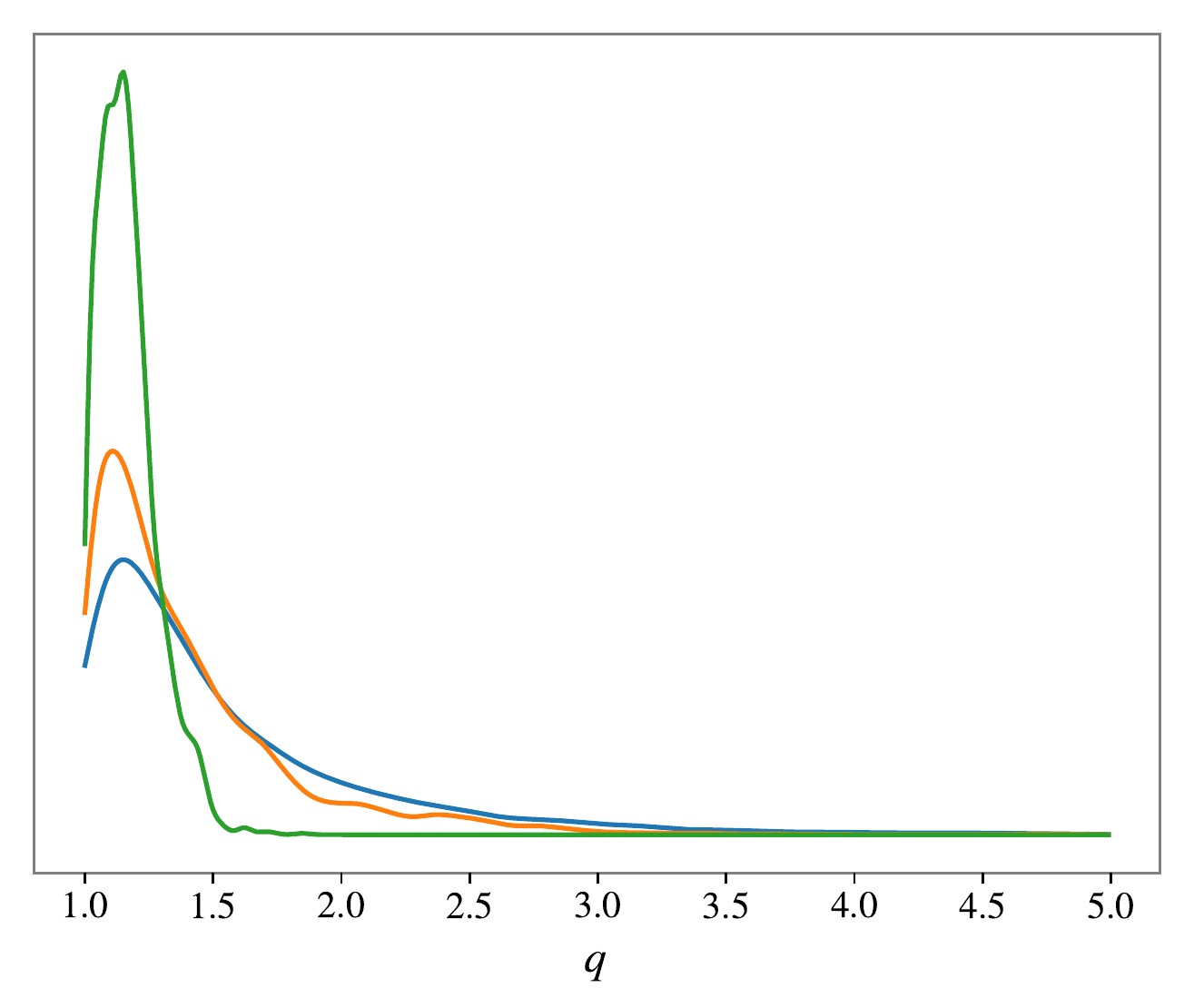}
\includegraphics[width=0.32\textwidth]{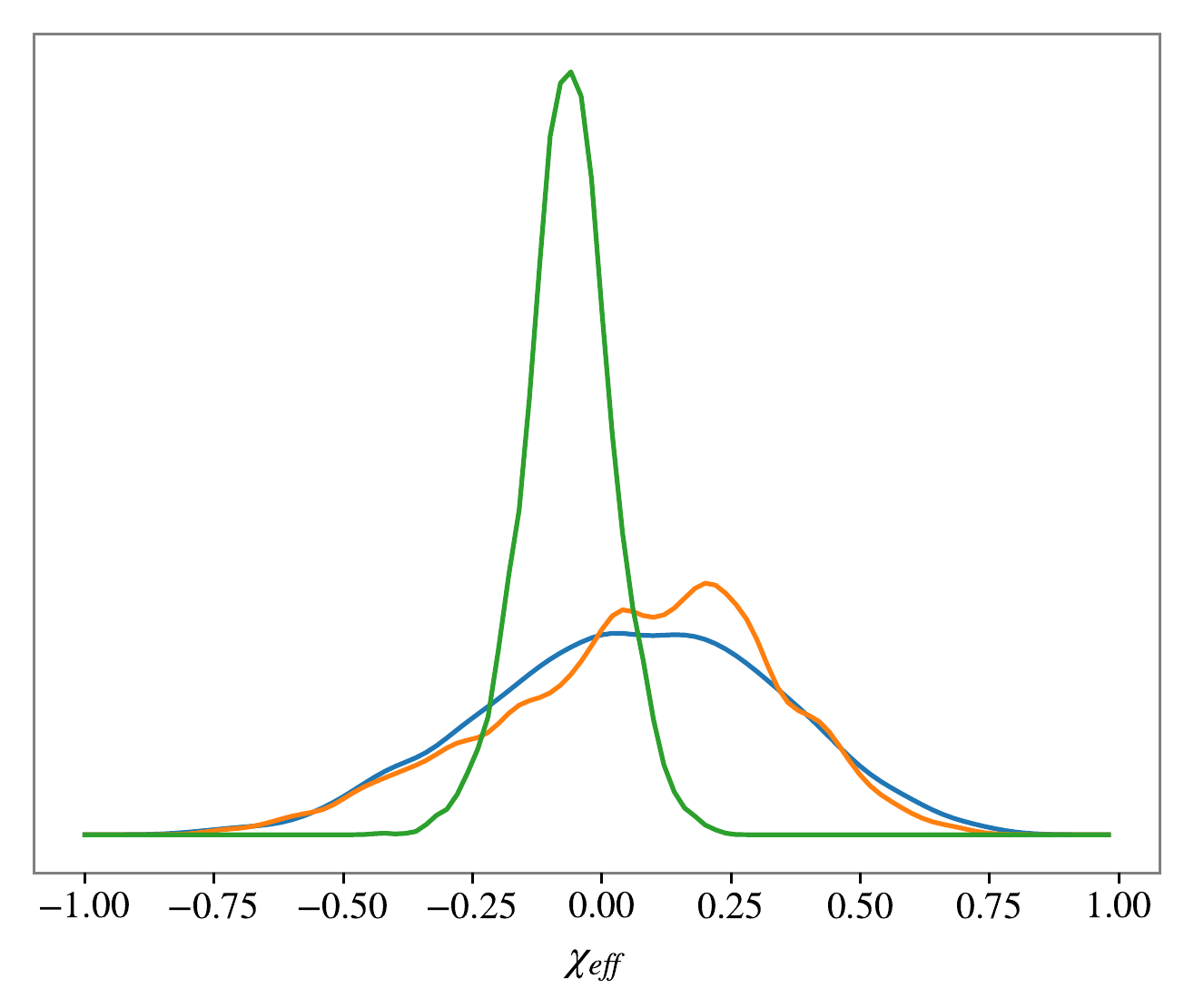}
\includegraphics[width=0.32\textwidth]{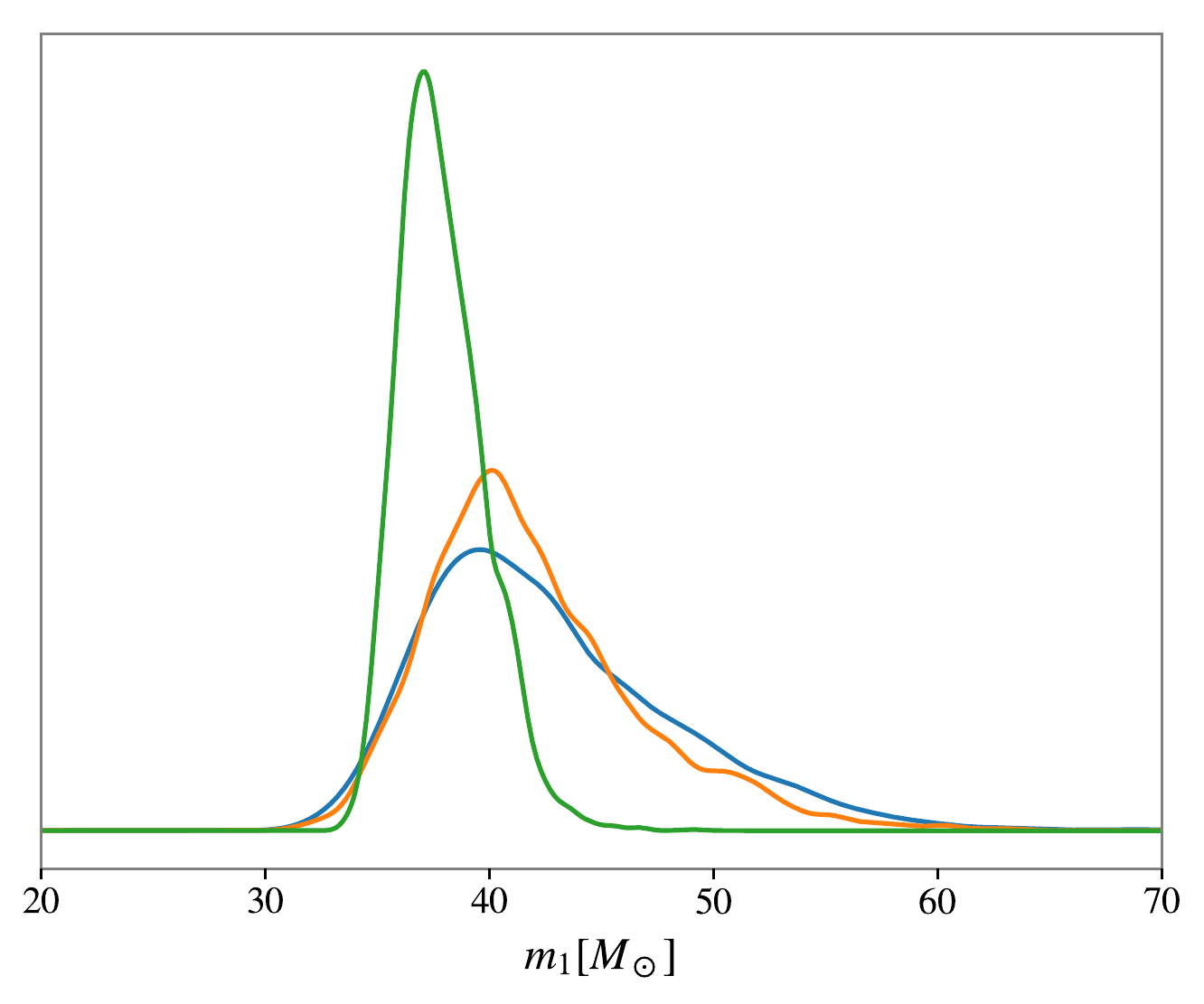}
\includegraphics[width=0.32\textwidth]{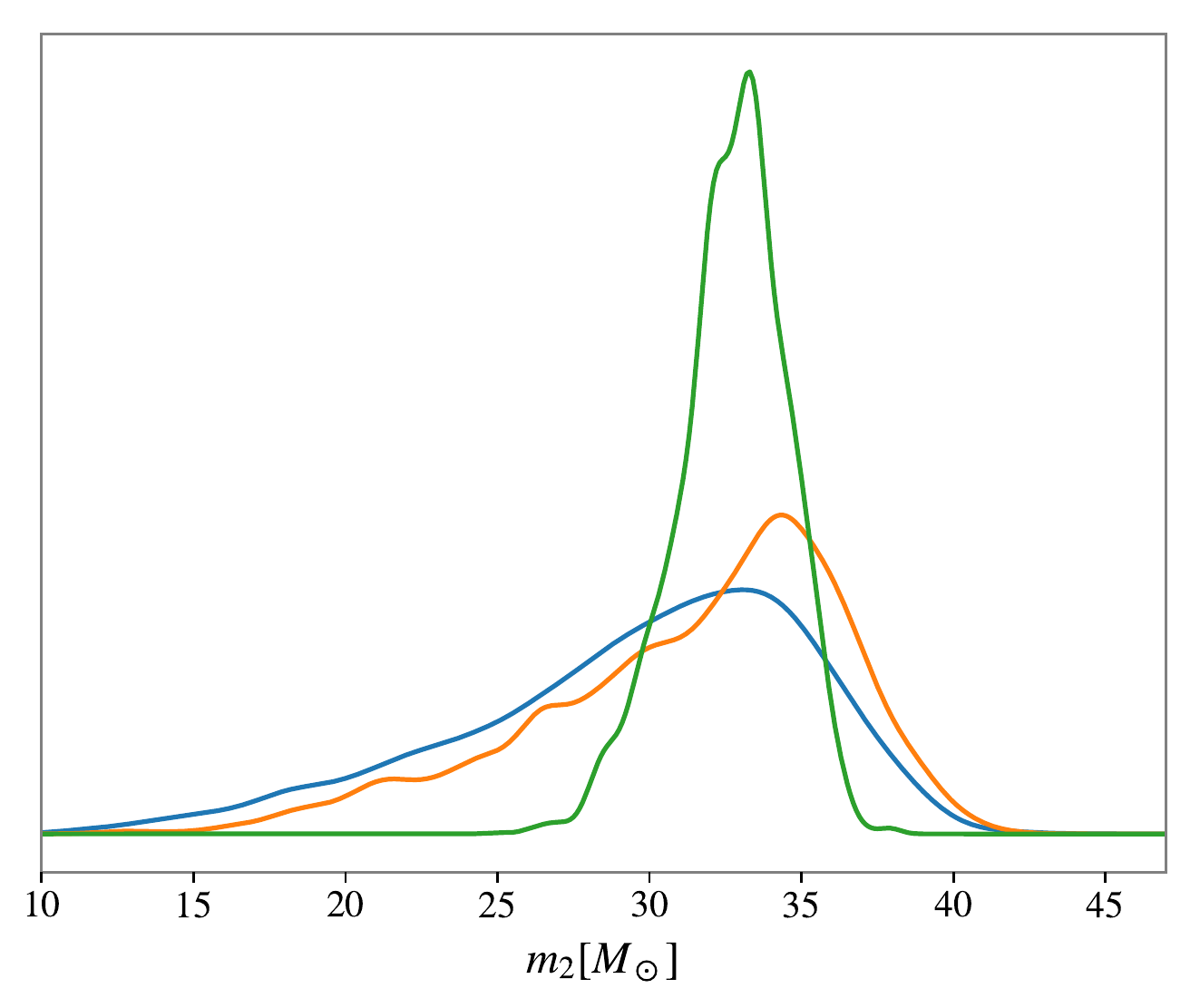}
\includegraphics[width=0.32\textwidth]{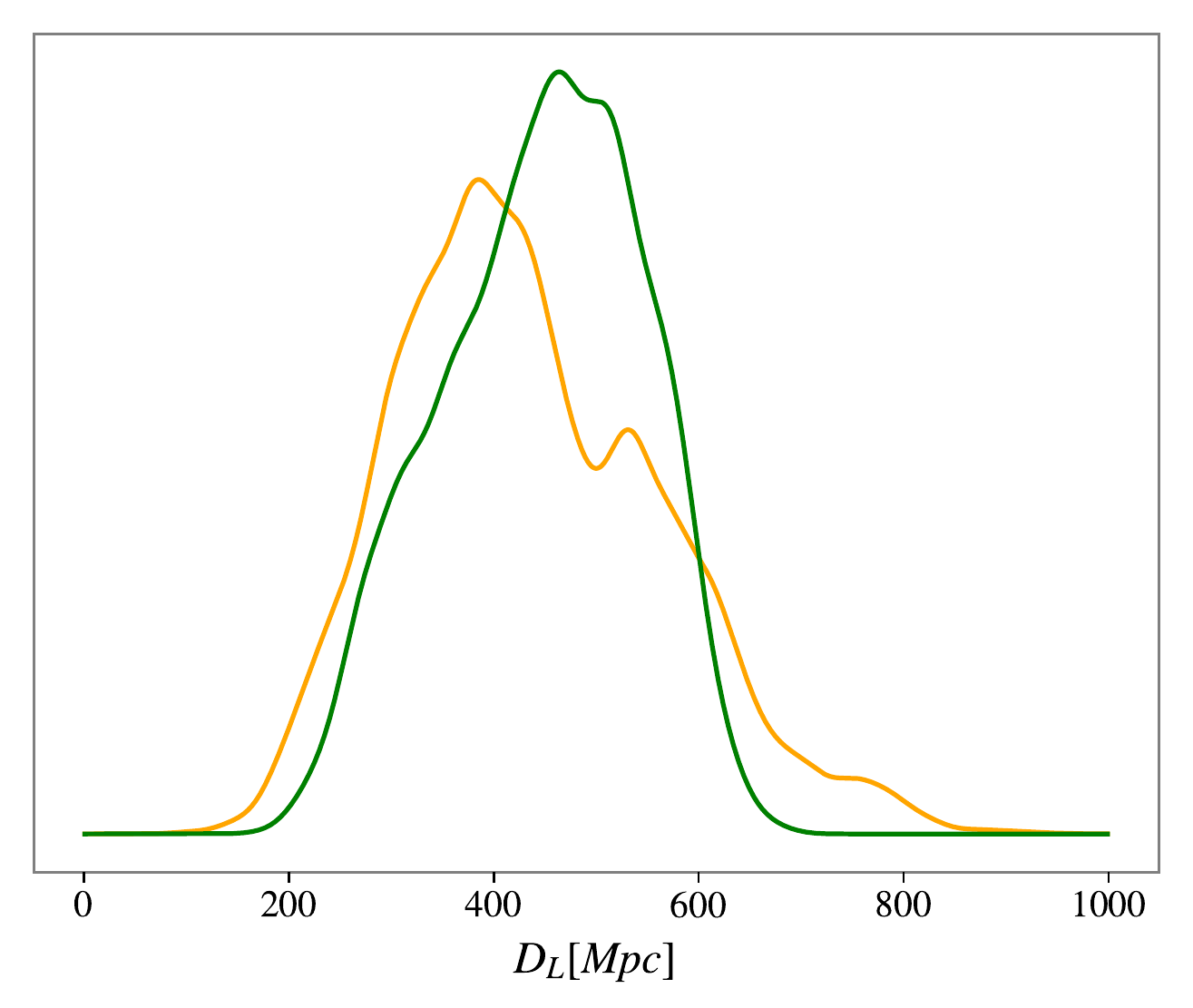}
\caption{\textbf{Binary parameters of GW150914, inferred from the post-peak emission}. Posterior distributions on individual masses, total mass, mass-ratio, effective spin parameter and luminosity distance obtained from the analysis of the post-peak signal of SXS:BBH:0305 (blue), the post-peak of GW150914 (orange) and the full GW150914 signal (green). All of the posteriors are consistent with one another other. Notably, we can constrain the mass ratio of GW510914 to $q<2.29$ at $90\%$ confidence analysing solely the post-peak portion of the signal, while the effective-spin parameter $\chi_{eff}$ can not be measured.} 
\label{fig:PhenomParameters}
\end{figure*}

The top-left panel of Figure \ref{fig:overtones} shows our posterior distributions for the  amplitude of the overtones relative to that of the fundamental tone for the case of our weaker injection and GW150914. In agreement with \cite{Giesler:2019uxc,Isi:2017fbj,Ota2020}, overtones exhibit amplitudes consistent with, or larger than that of the fundamental $(2,2,0)$ tone. For the case of GW150914, when including only one overtone, we obtain $A_1/A_0=1.37
^{+0.17}_{-0.20}$ at the $90\%$ credible level. When we use two, we obtain $A_1/A_0=1.10
^{+0.70}_{-0.90}$ and $A_2/A_0=0.70
^{+1.01}_{-0.55}$, showing that the signal residuals initially captured by the first overtone, eventually get similarly spread among both overtones. Similarly, for our weaker injection we obtain $A_1/A_0=0.96
^{+0.36}_{-0.46}$ when including one overtone and $A_1/A_0=0.87
^{+1.08}_{-0.79}$, $A_2/A_0=0.62
^{+0.90}_{-0.56}$ when including two. We stress once again, that while a value of zero may lay several $\sigma$ away from the center of the posterior distribution for the overtones, their presence is not strongly favoured by Bayesian model selection.
Finally, the bottom panels show that the estimated amplitude of the fundamental tone $A_0$ as successive overtones are added. We note that this always grows as more overtones are added. 

\subsection{Testing the no-hair theorem using overtones}
Here we focus on the evidence for the final object to be a Kerr Black hole. To this we compare the maximum evidence obtained by the Kerr models to that obtained by any of the Hair ones, irrespective of the number of modes. When only one mode is included in these two models, we obtain almost equal evidence. In general, as we add additional tones, we find that Hair models can fit the data better than the Kerr model using a lower number of tones. This is unsurprising; the Hair model has more flexibility to fit the data. As noted above, a better fit is not necessarily accompanied by a larger evidence. 
To understand this, it is best to first focus on the leftmost column of Table I, reporting our results for the injection with SNR = 100. As discussed earlier, the evidence for the Kerr model grows as we add up to two overtones. However, the tiny improvement of the fit obtained by the addition of a third overtone (or a fourth tone), makes the evidence decrease. The conclusion is that a scenario with two overtones (three tones) is preferred if the source is assumed to be a Kerr black hole.

For the Hair model, both the quality of the fit and the size of the parameter space grow faster as we include modes. For this reason, the evidence stops growing after the inclusion of just two tones, indicating that, if the source is not assumed to be a Kerr black hole, then the emission is best explained by only two tones. Comparing the maximum evidence for the Kerr and Hair models, we obtain $\log{\cal{B}}
^{Kerr, 2 overtone}_{Hair, 2 modes} \approx 2$, indicating that we cannot distinguish confidently between the scenarios of a Kerr black hole with three tones and an exotic object with 2 tones.

Reducing the SNR to 15 yields qualitatively similar results. We obtain $\log{\cal{B}}
^{Kerr, 0 overtone}_{Hair, 1 mode} \approx 0.3$ indicating, once again, that we are unable to distinguish between a Kerr black hole and an exotic object. For the case of GW150914, we also obtain consistent results, although in this case the presence of a second mode, in both the Kerr and Hair scenarios, is slightly preferred. We obtain $\log{\cal{B}}
^{Kerr, 1 overtone}_{Hair, 2 mode} \approx 0.6$, indicating that the Kerr scenario is only $1.8$ times more probable than the exotic object one.

\subsection{Testing the Kerr nature of the remnant using \texttt{IMRPhenomPv2}}
Last, we analyse the NR injection with the post-peak portion of \texttt{IMRPhenomPv2} waveforms. While we note that this model does only consider the $(2,2,n)$ modes of the emission, no evidence for higher-order modes has been reported for GW150914. This model allows us to effectively impose priors on the ringdown amplitudes and phases suitable for a perturbed Kerr black hole born from a binary black hole merger, effectively incorporating all overtones $n\leq\infty$. As expected, we find that this model is able to fit the data as well as the previous ones, while vastly reducing the parameter space. As a consequence, the binary black hole model is preferred with a $\log{\cal B}$ $\sim 5$ with respect to any of the other models for an SNR of 15. Performing this analysis on GW150914, we obtain a $\log{\cal B}$ of $\sim 6.5$ in favour of the binary black hole model, verifying the no-hair theorem at the $\sim 99.7\%$ level.

This result reveals that while no strong evidence can be obtained for the presence of one single overtone using our Kerr model, strong evidence for the whole set of overtones present in the signal can be obtained if we use the correct priors on their amplitudes and phases, even if it does not make sense to speak of measuring individual tones.

\begin{figure*}
\includegraphics[width=0.48\textwidth]{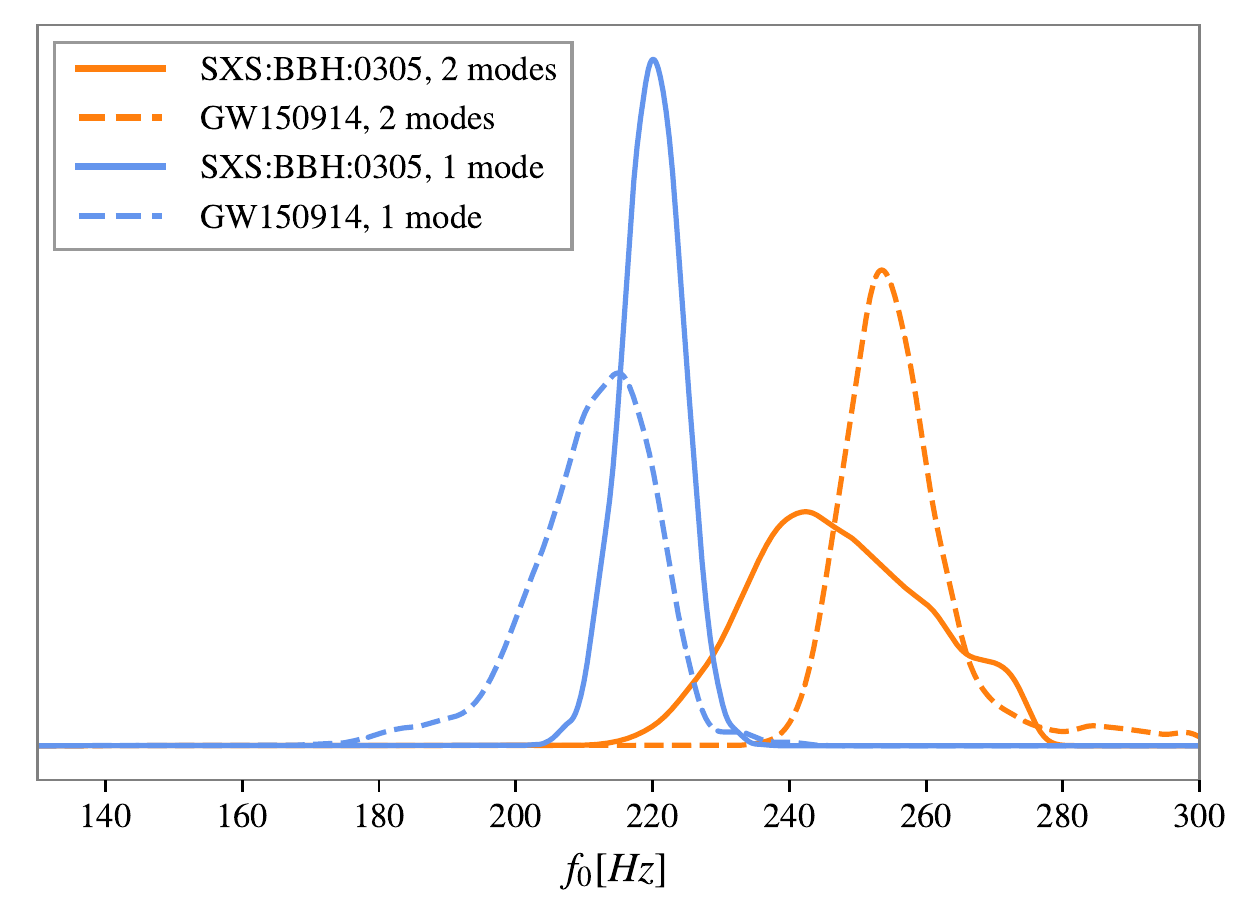}
\includegraphics[width=0.48\textwidth]{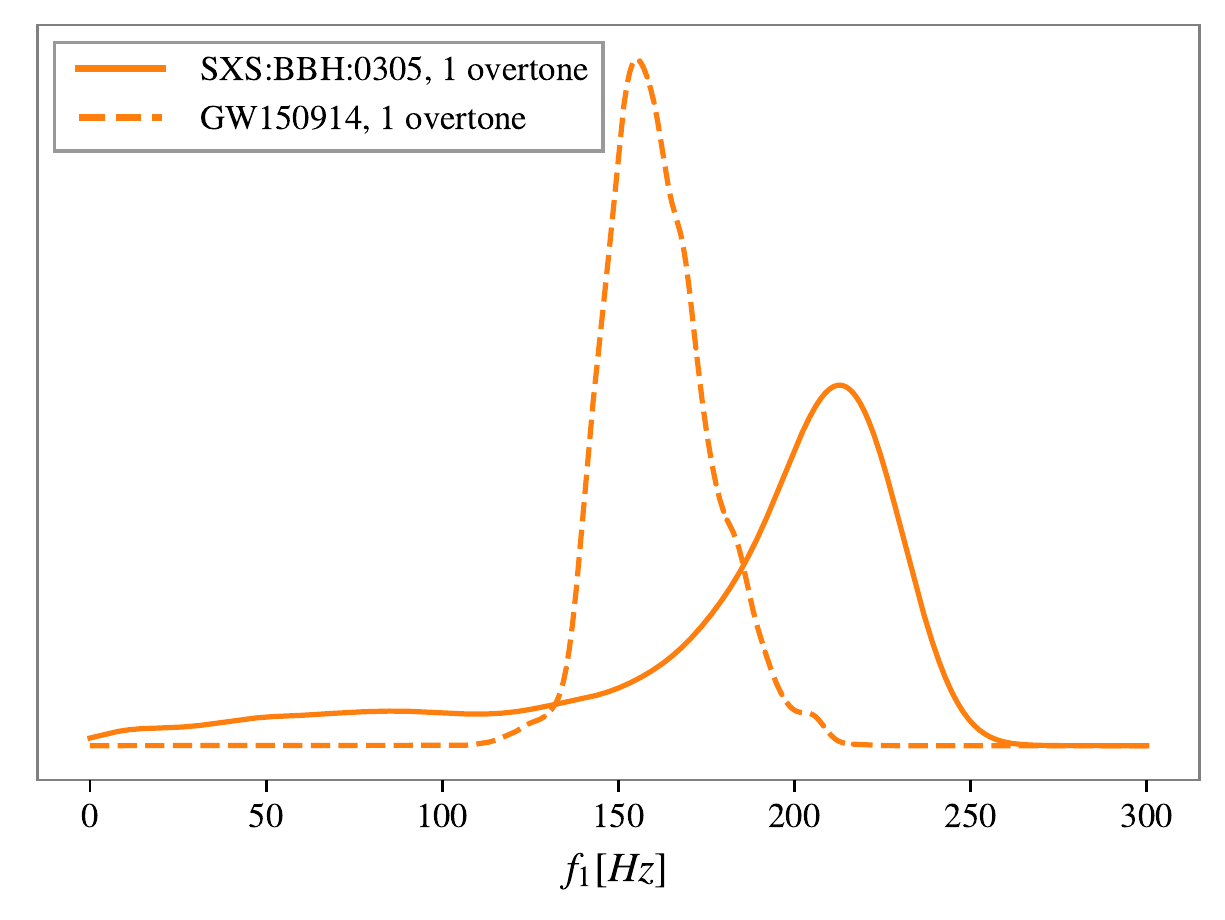}
\includegraphics[width=0.48\textwidth]{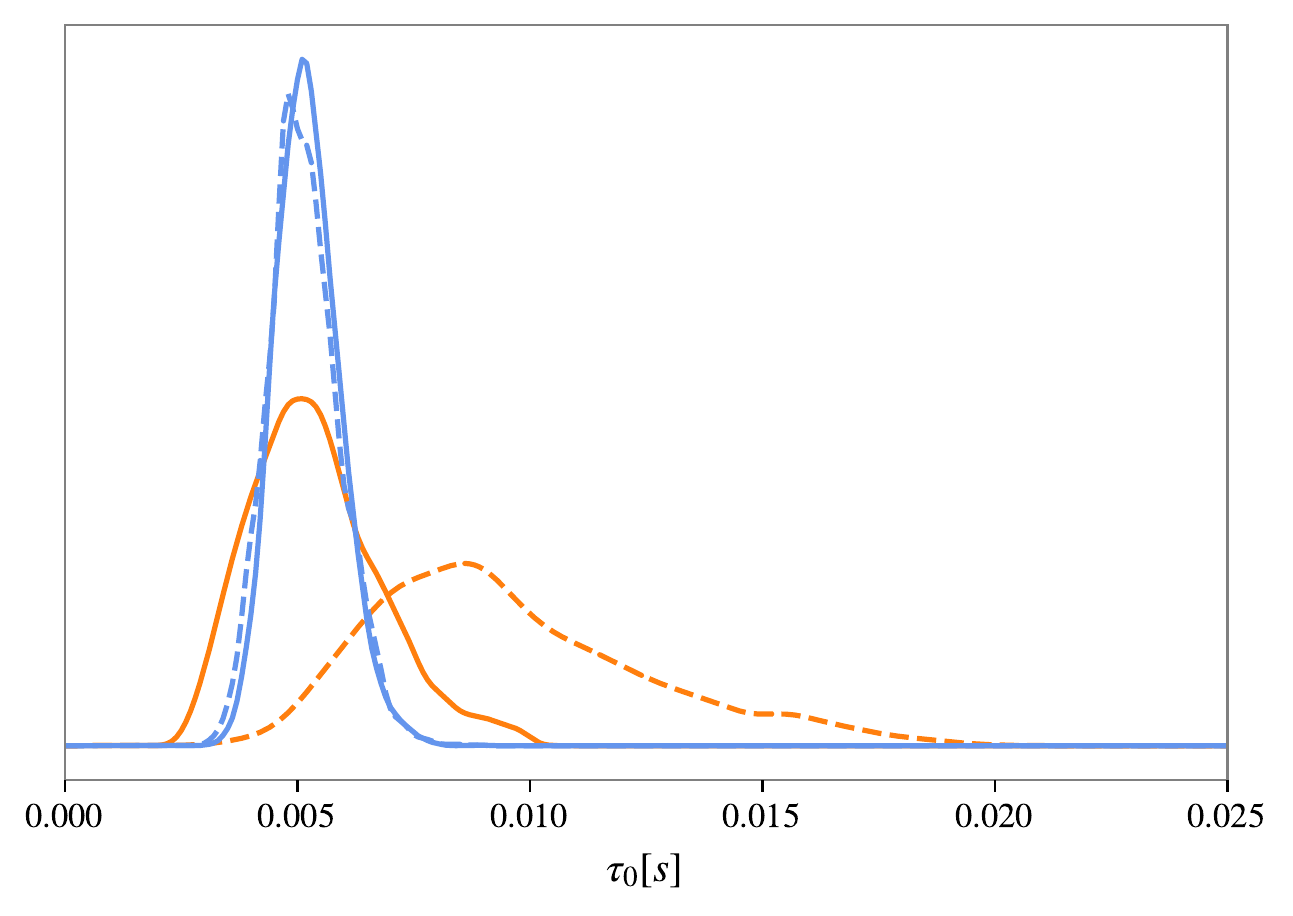}
\includegraphics[width=0.48\textwidth]{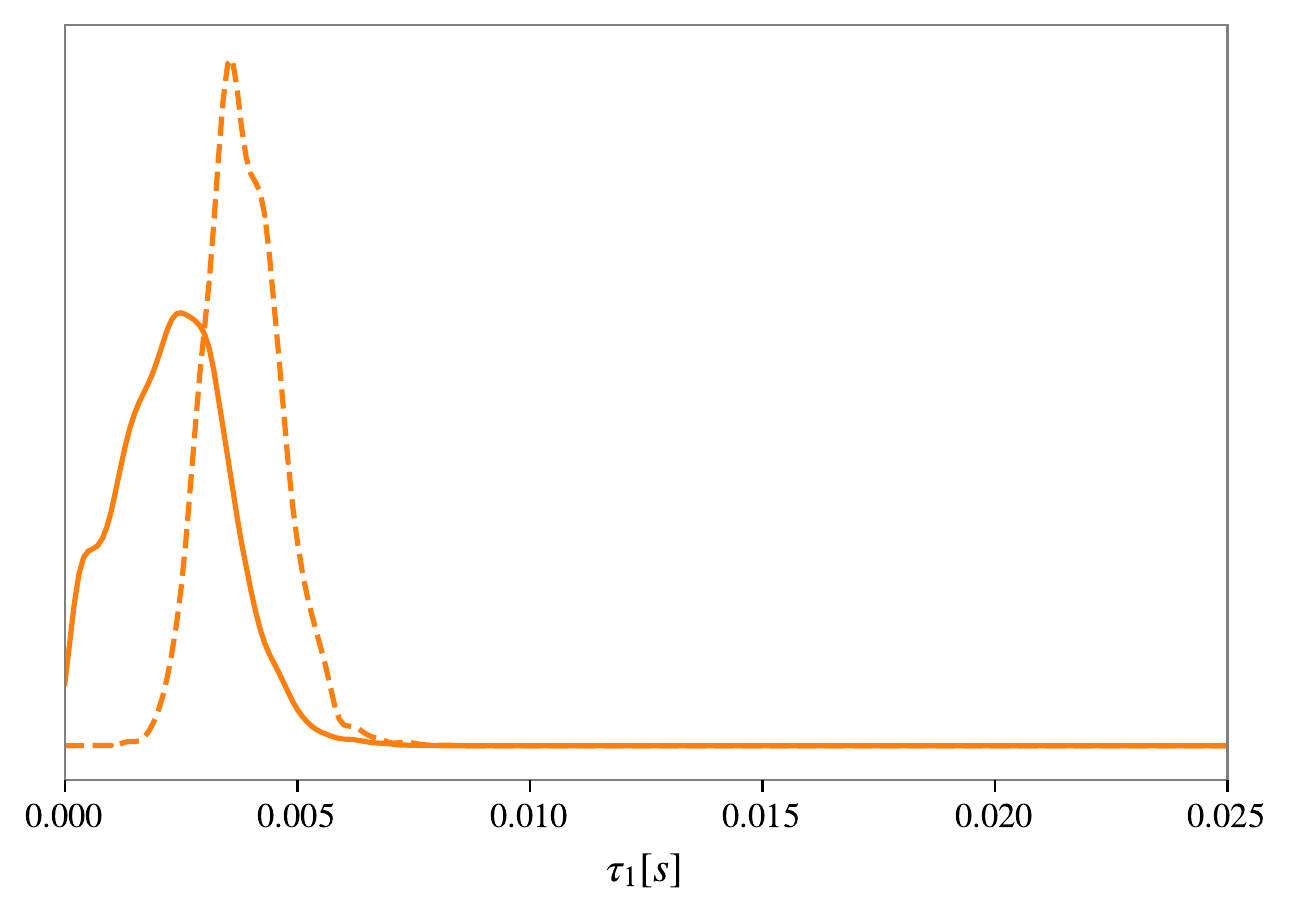}
\caption{Frequencies and damping times for GW150914 and a numerical injection consistent with it, with optimal SNR $\rho=15$. We report the frequency and damping time of the fundamental mode (left) and the next mode (right), when the signal is compared to one (blue) and two damped sinusoids (orange). When using two damped sinusoids, $f_0$ matches the ringdown frequency of the fundamental $(2,2,0)$ mode, while $f_1$ is consistent with the merger frequency. For GW150914, we find $f_0=254^{+6}_{-7}$Hz, consistent with (TGR paper).  For both the injection and the GW150914 case, $\tau_0$ shows significant support for values larger than that of the $(2,2,0)$ mode. The value of $f_1$ obtained in both cases is consistent with the merger frequency. For GW150914 we obtain $f_1=150^{+16}_{-12}$Hz.} 
\label{fig:hair}
\end{figure*}

\subsection{Recovering the binary parameters from the post-merger signal}
By imposing suitable priors on the amplitudes and phases of the overtones, we recover information about the parent binary encoded in the infinite series of overtones, leading, in turn, to a large preference for \texttt{IMRPhenomPv2}. 
Figure~\ref{fig:PhenomParameters} shows the posterior distributions for the binary parameters inferred from the post-peak of the NR simulation scaled to $\rho=15$ and that of GW150914; together with those obtained from the analysis of the full GW150914 signal. All these results are in agreement. We find that information about the parent binary masses can be recovered solely from the post-merger signal, although with larger uncertainties that when analysing the full inspiral-merger-ringdown signal. We obtain $90\%$ credible intervals for the individual (redshifted) masses $m_1=41
^{+10}_{-5}M_\odot$ and $m_2=32
^{+6}_{-11}M_\odot$, constraining the mass ratio of the parent binary to $q<2.29$ at the $90\%$ level. We find that relatively little information about the binary spins, encoded in the $\chi_{eff}$ parameter, is recovered using data from after the signal peak, as both this posterior and those for the individual spin magnitudes are perfectly consistent with our priors. The same remains true when we allow for precessing spins. Consistently, Table I reports almost identical evidences for all of the non-spinning, aligned-spin and precessing-spin models.\\

Finally, we note that while this work was being performed, Jim\'{e}nez et. al., released a model for the remnant of non-spinning BBHs parametrising both the fundamental $(2,2,0)$ mode and its first overtone as a function of the binary parameters \cite{Xisco}. However, we analysed our injection with SNR$=15$ with this model obtaining biased results. In particular, while we obtain one-dimensional $90\%$ credible intervals for the mass-ratio and total mass of $q = 1.70^{+2.0}_{-0.6}$ and $M_f = 78.5^{+4.5}_{-8.5}M_\odot$ that include the true injection values, namely $q=1.36$ and $M=72M_\odot$, these lay way out of the two-dimensional $90\%$ credible region. We understand that this is consistent with the fact that \cite{Xisco} reports that their model has mismatches as large as 0.02 with numerical relativity waveforms for mass ratios close to unity.

\subsection{Reconstructing the ringdown waveform with minimal assumptions}
Last, we perform spectroscopic parameter estimation with minimal assumptions.
We use the Hair model consisting of damped sinusoids with free frequencies and damping times. 
Figure~\ref{fig:hair} shows posteriors for the damping times (bottom) and frequencies (top). Analysing our NR simulation with a single tone yields $90\%$ credible intervals $f_0 =220^{+7}_{-7}$Hz and $\tau_0=5.1^{+1.3}_{-1.0}$ ms consistent with those reported in Thrane et. al. \cite{Thrane:2017lqn}. A consistent result $f_0 =213^{+11}_{-17}$Hz, $\tau_0=5.1^{+1.3}_{-1.1}$ is obtained for GW150914. Adding a second tone, we estimate the frequency of the zeroth mode to be $f_0=254
^{+21}_{-9}$Hz, consistent with the value for the fundamental $(2,2,0)$ Kerr mode $f_0=251
^{+8}_{-8}$Hz reported in \cite{TheLIGOScientific:2016src}. For the second tone we estimate $f_1=160^{+28}_{-19}$ Hz, consistent with that at the merger of GW150914 \cite{Abbott:2017xlt}. This result is sensible: we resolve the frequency of the longest-lived tone and that at the amplitude peak \cite{Abbott:2017xlt}. For the NR simulation, we obtain consistent values $f_0=247
^{+23}_{-19}$Hz and $f_1=201^{+31}_{-140}$ Hz.

\section{\textbf{Conclusions}} 
Black-hole spectroscopy pursues the measurement of  different gravitational-wave modes emitted during black-hole ringdowns. According to the no-hair theorem, the frequencies and damping times of these modes are fully determined by the black hole mass and spin while the amplitudes and phases depend on the particular perturbation it undergoes. Consequently, measuring at least two of these modes seems key to confirm the theorem.

Performing Bayesian model selection on simulated signals consistent with GW150914, we assess the feasibility of tests of the no-hair theorem based on the individual identification of ringdown overtones. The large number of degrees of freedom needed to correctly fit the post-merger signals from black holes, together with the flexibility that models violating the no-hair theorem offer, makes it impossible to distinguish between two possible scenarios: a) a generically perturbed black-hole with quasinormal modes that satisfy the no-hair theorem and b) an exotic object with fewer active quasinormal modes, which violates the no-hair theorem. This is true even for loud signals. The reason is that, the louder the signal, the larger the number of overtones is needed to correctly fit it, which increases the number of required parameters.

In this work, we highlight the need to place suitable priors on these parameters by exploiting the information that the remnant black hole is the result of a binary merger, so that the number of free ringdown parameters should never exceed eight. Instead of attempting to measure individual overtones, we employ the post-merger portion of complete inspiral-merger-ringdown waveform that naturally include all the possible overtones.
These waveforms are parametrised by solely the binary parameters. Analysing a numerical simulation consistent with GW150914, we show that this reduction of the parameter space allows us validate the no-hair theorem with strong confidence. 
Applying this method to GW150914, we obtain a natural log Bayes factor of $\approx 6.5$ favouring the Kerr nature of the remnant object, and its compliance with the no-hair theorem. This leaves 1 in $\sim 600$ chances that the theorem is violated by GW150914. Interestingly, using only the post-merger emission, we can constrain the mass ratio of the parent binary to $q \leq 2.29$ at the $90\%$ level.
By measuring a property of the progenitor system (as opposed to a property of the remnant), it is clear that the amplitudes and phases of the overtones encode detailed information about how the remnant black hole was perturbed. 


We conclude that the most meaningful way to test the no-hair theorem is the direct comparison with full waveform models, in contrast to performing minimal-assumption black-hole spectroscopy. In the future, we plan to extend this study to the case of angular ringdown modes.

\section{\textbf{Acknowledgements}}
It is our pleasure to thank Gregorio Carullo, Miriam Cabero and Tjonnie Li for useful discussions and detailed feedback. We also thank Colm Talbot and Ethan Payne for their help with the parameter estimation code \textsc{Bilby}. We acknowledge the support of the Australian Research Council grants DP180103155, CE170100004, FT160100112, and FT150100281. JCB is partially supported by the Direct Grant, Project 4053406, from the Research Committee of the Chinese University of Hong Kong. The project that gave rise to these results also received the support of a fellowship from ”la Caixa” Foundation (ID
100010434) and from the European Union’s Horizon
2020 research and innovation programme under the
Marie Skłodowska-Curie grant agreement No 847648.
The fellowship code is LCF/BQ/PI20/11760016. The authors acknowledge computational resources provided by the LIGO Laboratory and supported by
National Science Foundation Grants PHY-0757058 and PHY0823459; and the support of the NSF CIT cluster for the provision of computational resources for our parameter inference runs. This document has LIGO DCC number P-2000372. 

\section*{Appendix I: Bayesian Inference and Occam Factor}

\subsection{Basic definitions}

The posterior probability for a set of source parameters $\theta$, given a stretch of data $d$ and a data model ${\cal{M}}$, is given by \begin{equation}
p(\theta| d,  {\cal{M}}) = \frac{\pi(\theta) {{\cal{L}}(d|\theta,{\cal{M}})} }{{\cal{Z}}(d|{\cal{M}})},
\label{eq:bayes}
\end{equation}
where ${\cal{ L}}$ denotes the standard frequency-domain likelihood commonly used for gravitational-wave transients \cite{Finn1992,Romano2017}
\begin{equation}
 \log{\cal{L}}(d|\theta,{\cal{M}}) =  - \frac{1}{2}(d-h_{{\cal{M}}}(\theta)|d-h_{{\cal{M}}}(\theta)).
\end{equation}
Here, $h_{\cal{M}}(\theta)$ denotes a waveform template for parameters $\theta$, according to the waveform model ${\cal{M}}$. As usual, the operation $(a|b)$ denotes the inner product \cite{Cutler:1994ys}
\begin{equation}
    (a|b)= 4 \Re \int_{f_{\min}}^{f_{\max}} \frac{\tilde{a}(f)\tilde{b}(f)}{S_n(f)} df,
\end{equation}
where $S_n(f)$ denotes the one sided power spectral density of the detector noise, and $f_{\min}$ and $f_{\max}$ are respectively the low and high frequency cutoffs of the detector data. The factor $\pi(\theta)$ denotes the prior probability for the parameters $\theta$ and the factor ${\cal{Z}}(d|{\cal{M}})$ is known as the evidence for the model ${\cal{M}}$. This is given by the integral of the numerator of Eq.\ref{eq:bayes} across all the parameter space covered by the model
\begin{equation}
{\cal{Z}}(d|{\cal{M}}) := {{\cal{Z}}}_{{\cal{M}}} = \int \pi(\theta) {{\cal{L}}(d|\theta,{\cal{M}})} d\theta.
 \label{eq:z}
\end{equation}

Given two models $A$ and $B$, the degree of preference for model $A$ over model $B$ is given by the Bayes' Factor
\begin{equation}
{\cal{B}}^{A}_{B} = \frac{{\cal{Z}}_A}{{\cal{Z}}_B}.
\end{equation}
It is common to say that model $A$ is strongly preferred wrt. $B$ when the natural log${\cal{B}}^{A}_{B} > 5$, so that model A is $\sim 150$ times more probable than model B. \\

\subsection{The size of the parameter space and the Occam factor}

When comparing two models $A$ and $B$, it is important to note that two main factors determine the value of the corresponding evidences. The first one is how well the model can fit the data. Parameters yielding good fits will yield large values of ${\log{\cal{L}}}$, and vice versa. In particular, note that ${\cal{Z}}$ is bounded by, e.g.,
\begin{equation}
{\cal{Z}} \leq \int \pi(\theta) {\cal{L}}_{Max} d\theta, 
\end{equation}
with $\log{\cal{L}}_{Max}$ denoting the maximum value of the likelihood across the parameter space, i.e., at the best fitting parameters. Second, the act of integrating across the whole parameter space implies that the model may explore regions of the parameter space with poor contributions to the integral. Since $\int\pi(\theta) d\theta = 1$ 
exploring ``useless'' portions of the parameter space leading to poor fits causes a reduction of ${\cal{Z}}$. This penalty is known as the \textit{Occam factor}. Three situations can arise when adding a new parameter $\theta_1$ to an existing set of parameters $\theta$.

\subsubsection{The new parameter $\theta_1$ has no effect on the fit}In this case, we have that ${\cal{L}}(d|\theta) = {\cal{L}}(d|\theta,\theta_1)$. For instance, when looking solely at the post-merger of GW150914, the spins of the parent binary will not have any appreciable effect on the signal, so that the spin values will not have any effect on the likelihood. Fig. 3 shows this is the case for us, as the prior and posterior $\chi_{eff}$ distributions are hardly different, indicating that the data is not informative about the spins. Accordingly, Table I shows no preference for either model.

\subsubsection{The new parameter $\theta_1$ significantly improves the fit}
This is, if ${\cal{L}}(d|\theta) \neq {\cal{L}}(d|\theta,\theta_1)$ and ${\cal{L}}_{max}(d|\theta,\theta_1) >> {\cal{L}}_{max}(d|\theta)$, roughly speaking, the average value of ${\cal{L}}$ across the parameter space will increase.
An example of this can be observed for our injection with SNR = 100. The addition of a second mode to the templates causes a very important increment of the maximum likelihood that overcomes the fact of exploring extra poor matching parameter combinations. As a consequence, the first column of Table I shows that the Bayes factor significantly increases when going from one single mode to two. As we have discussed, this also happens when the full tower of overtones is added to the fundamental mode, with its parameters suitably constrained, i.e., when using the \texttt{IMRPhenomPv2} model.

\subsubsection{The new parameter does not significantly improve the fit} Consider that the new parameter only leads to a marginal improvement of the fit at the best point in the parameter space, so that ${\cal{L}}_{max}(d|\theta,\theta_1) \approx {\cal{L}}_{max}(d|\theta)$. In addition, consider that $\theta_1$ leads to poor fits most of the times, so that roughly speaking, the average value of the likelihood across the new space decreases. In these case, the marginal improvement of ${\cal{L}}$ max does not compensate the increase of the parameter space. A good example of this is the addition of a third and a fourth mode to the Hair model, in the first column of Table I. A similar situation arises when adding a third overtone to the Kerr model.

\section*{Appendix II: choice of priors}
The selection of prior ranges is a fundamental ingredient of Bayesian inference and model selection that often requires a very careful analysis to avoid biasing the analysis towards one of the compared models. The simplest situation occurs when the two compared models ${\cal{A}}$ and ${\cal{B}}$ are so that one is contained in the other (i.e., the models are nested), and the simpler model is described by a subset of the parameters describing the more complex one. For instance, evidence in favour of the presence of non-zero spins in a binary black hole can be evaluated by comparing the data to two models that include and omit spin effects while covering the same prior ranges in all other parameters. This way, we know that any difference in the evidence for these models comes solely from the impact of the black hole spins in the data.

The situation, however, becomes more involved when the compared models do not depend on the same parameters, as it is our case.
With the exception of the orientation and sky-location parameters, our models do not depend explicitly on the same parameters, which makes less trivial to set appropriate priors. For the Hair model, the intrinsic properties of the source are described in terms of the individual amplitudes, phases, damping times and frequencies of the emission modes $\{A_n,\phi_n,f_n,\tau_n\}$. Instead, for the Kerr model, the $\{f_n,\tau_n\}$ parameters are parametrised by the mass and spin of the black hole. Finally, the model \texttt{IMRPhenomP} effectively parametrises all $\{A_n,\phi_n,f_n,\tau_n\}$ via the individual masses and spins of the binary black hole. For this reason, choosing priors that do not artificially favor either of these models is in principle a delicate matter. For instance, a prior too wide in the $\{\tau_n,f_n\}$ parameters for the Hair model, together with a too narrow prior on $\{M_f,a_f\}$ for the Kerr model, would artificially favor the former. In the following, we describe our prior choices.

\subsubsection{Hair vs. Kerr}

\begin{enumerate}
\item We set flat priors in all mode phases $\phi_n \in [0,2\pi]$ and a flat prior in the amplitude of the zeroth mode $A_0 \in [10^{-22},10^{-18}]$. 

\item In the Kerr case, we set the amplitude of the overtones to $A_i \in [10^{-23},10^{-18}]$. For the secondary $n>0$ modes we apply priors $A_i/A_0 \in [0.01,100]$. While this choice makes sampling easier, leading to cleaner posterior distributions, we have checked that our conclusions do not change if we use a prior $A_i \in [10^{-23},10^{-18}]$. In addition, we apply an extra constraint $f_n / f_{n-1} \in [0.01, 1]$ and $\tau_n /\tau_{n-1} \in [0.01, 1]$. This is motivated by the fact that, in the Kerr case, overtones have lower frequencies and damping times than the fundamental mode. 

\item We set flat priors on the mass and spin of the Kerr model $M_f \in [20,100]M_\odot$ and $a_f \in [0,0.99]$.
\item We set flat priors on the frequency and damping time of the fundamental Hair mode, $f_0$ and $\tau_0$, equal to the maximum and minimum explored by the masses and spins of the Kerr model.
\end{enumerate}

\subsubsection{IMRPhenomP}

For $\texttt{IMRPhenomP}$ we set a flat prior in the total mass $M \in [20, 100] M_\odot$ and uniform priors in the spin magnitude $a_i \in [0.,0.8]$. The luminosity distance is allowed to vary in $d_L \in [1,1000]$, so that it runs over the same orders of magnitude as $A_0$ in the previous two models.


\bibliography{IMBBH.bib}

\end{document}